\documentclass[12pt]{amsart}

\usepackage[margin=2.5cm]{geometry}
\usepackage{tikz}
\tikzset{node distance=2cm, auto}
\usetikzlibrary{matrix,arrows,decorations.pathmorphing}

\usepackage{graphicx}
\usepackage{amsmath,amssymb}

\usepackage{amscd}
\usepackage{verbatim}
\usepackage{enumerate}

\newtheorem{theorem}{Theorem}

\newtheorem{proposition}{Proposition}

\newtheorem{remark}{Remark}

\newcommand{\g}{\mathfrak{g}}

\newcommand{\hh}{\mathfrak{h}}
\newcommand{\kf}{\mathfrak{k}}

\newcommand{\RR}{\mathbb{R}}
\newcommand{\CC}{\mathbb{C}}

\newcommand{\pa}{\partial}

\newcommand{\tr}{\operatorname{tr}}

\newcommand{\End}{\operatorname{End}}

\newcommand{\be}{\begin{equation}}
\newcommand{\ee}{\end{equation}}

\newcommand{\cM}{\mathcal{M}}
\newcommand{\cS}{\mathcal{S}}

\newcommand{\cO}{\mathcal O}
\newcommand{\cP}{\mathcal P}

\newcommand{\cB}{\mathcal B}
\newcommand{\bS}{{\bf S}}

\usepackage[colorlinks=true]{hyperref}
\usepackage{stmaryrd} 
\usepackage{commath} 
\usepackage{empheq}  

\newcommand{\inv}{^{-1}}
\newcommand{\tensor}{\otimes}
\newcommand{\hf}{\mathfrak h}

\DeclareMathOperator{\Ad}{Ad}

\newenvironment{dedication}
        {\vspace{6ex}\begin{quotation}\begin{center}\begin{em}}
        {\par\end{em}\end{center}\end{quotation}}

\title{Spin Calogero-Moser models on symmetric spaces}

\author{Nicolai Reshetikhin}
\address{N.R.: Department of Mathematics, University of California, Berkeley,
CA 94720, USA \& Physics Department, St. Petersburg University, Russia \&KdV Institute for Mathematics, University of Amsterdam,
Science Park 904, 1098 XH Amsterdam, The Netherlands.}
\email{reshetik@math.berkeley.edu}

\begin{document}

\maketitle

\begin{dedication}
\hspace{4cm}
\vspace*{3cm}{Dedicated to the memory of B.A.Dubrovin.}
\end{dedication}

\begin{abstract}
In this paper we construct and prove superintegrability of spin Calogero-Moser type systems 
on symplectic leaves of $K_1\backslash T^*G/K_2$ where $K_1,K_2\subset G$ are 
subgroups. We call them two-sided spin Calogero-Moser systems. 
One important type of such systems correspond to $K_1=K_2=K$ where $K$ is a subgroup of fixed points of Chevalley
involution $\theta: G\to G$. The other important series of examples come from pair $G\subset G\times G$
with the diagonal embedding.  We explicitly describe examples of such systems corresponding to symplectic leaves
of rank one when $G=SL_n$.

\end{abstract}

\section*{Introduction}

\subsection*{1}Calogero-Moser models are integrable Hamiltonian systems describing 
interacting one-dimensional many particle systems\footnote{Calogero and Moser introduced systems with rational potential corresponding to the root system
of type $A_n$. Sutherland \cite{Suth} generalized them to trigonometric potentials. Olshanetsky and Perelomov extended these models to arbitrary root systems,
see \cite{OP} for an overview of Calogory-Moser systems. See also the collection \cite{CMbook} for more recent developments.} \cite{Ca}\cite{Mo}.

Spin Calogero-Moser systems appeared as generalizations of these models involving extra degrees of freedom which are usually interpreted as "spin variables" \cite{GH} and were initially called Euler-Calogero-Moser systems since they can be considered as 
an Euler top interacting with a Calogero-Moser system. Integrability of spin Calogero-Moser systems was studied in numerous papers from various points of view, see for example \cite{KBBT}\cite{CMbook}\cite{LX}.
The superintegrability\footnote{The notion of superintegrability is reviewed below. It is also known as degenerate integrability \cite{N} or non-commutative integrability.}  of this model was proven in \cite{R1}, see also references therein. The phase space of a spin Calogero-Moser system is a symplectic leaf of the Poisson space $T^*G/G$, 
where the action of $G$ is the natural lift to $T^*G$ of the conjugation action of $G$
on itself.  The Poisson commuting Hamiltonians are $Ad_G$-invariant functions on $\g^*$ pulled back through the projection $T^*G/G\simeq (\g^*\times G)/G\to \g^*/G$. Here we trivialized the cotangent bundle $T^*G\simeq \g^*\times G$ by right translations (or, equivalently, by left translations). The space $T^*G/G$ is singular and its symplectic leaves are singular. 
More precisely  they are stratified symplectic spaces \cite{SL}. However, it has an open dense (Zariski open in the algebraic setting) subspace which is a smooth algebraic orbifold $T^*G_{reg}/G$, where $G_{reg}$ is a subset of regular elements in $G$ . 
Its symplectic  leaves are largest symplectic strata of symplectic leaves of  $T^*G/G$. Poisson structure on $T^*G/G$ means a Poisson bivector field in a vicinity of smooth points of $T^*G$.

First examples of spin Calogero-Moser systems corresponding to symmetric pairs $K\subset G$ appeared in  \cite{Fe} (see also references therein). Such systems  have symplectic leaves of  $T^*(G/K)/K$ as theirs phase space.  Here the action of $K$ on $T^*(G/K)/K$ is the natural extension of the left action of $K\subset G$ on $G$. These are Hamiltonian systems that have Hamiltonians which are pull-backs of $Ad^*$-invariant functions on $\g^*$ with respect to the natural
projection $T^*G\to \g^*$. Throughout this paper we assume the trivialization of cotangent bundle $T^*G\simeq \g^*\times G$ by right translations. The number of such independent Hamiltonians is the rank of $G$ and is, typically,
less than half of the dimension of the phase space. 

On the other hand, the equations of motion in such systems
can be solved by the method of Hamiltonian reduction \cite{Ko}, now also known as the factorization method \cite{STS}.
By analogy with characteristic Hamiltonian systems on simple Lie groups \cite{R} and with the spin Calogero-Moser systems \cite{R1}, one can expect that they are superintegrable\footnote{See \cite{SInt}\cite{N}\cite{R2} for a history and more details on superintegrability} in a sense of \cite{N}. Here we will show that two--sided spin Calogero-Moser systems are indeed superintegrable.

\subsection*{2} Now recall the notion of superintegrability in Hamiltonian mechanics \cite{N}\cite{SInt}. Let $(\cM,\omega)$ be a symplectic manifold\footnote{ We assume
that $\cM$ is a smooth manifold. But it can also be a complex holomorphic manifold or an algebraic variety. All these cases are relevant for spin Calogero-Moser systems which are studied in this paper. Note that phase spaces of Calogero-Moser systems were studied in algebro-geometric setting extensively, see for example \cite{CMspaces}. } 
of dimension $2n$. 

A {\it superintegrable system} on $\cM$ is a pair of Poisson projections $p_1$ and $p_2$
\begin{equation}\label{SI}
\cM\stackrel{p_1}\rightarrow \cP \stackrel{p_2}\rightarrow\cB
\end{equation}
where $\cP$ is a Poisson manifold of dimension $2n-k$, $\cB$ is a $k$-dimensional  manifold with the trivial Poisson structure, and fibers of $p_2$ are symplectic leaves in $\cP$. Let $C(\cM)$ be the algebra of functions
on $\cM$ (smooth if $\cM$ is a smooth manifold etc.). Let us identify $C(\cB)$ and $C(\cP)$ with
subalgebras in $C(\cM)$. Then the above properties of projections $p_1$ and $p_2$ are
equivalent to the property that Poisson centralizer
of $C(\cB)\subset C(\cM)$ is $C(\cP)$. 

In the superintegrable systems, functions from $C(\cB)$ are Poisson commuting Hamiltonians and
functions from $C(\cP)$ are integrals of motions, which do not necessary commute.
Generic level sets of integrals of motions are isotropic submanifolds in $\cM$.
Flow lines of any $k$ independent functions on $\cB$ provide an affine coordinate 
system on such level sets and the flow line of any function on $\cB$ is linear 
in these coordinates. For more details on superintegrable systems see \cite{N}\cite{SInt}\cite{R2}.

\subsection*{3}In this paper we will construct Calogero-Moser type
Hamiltonian systems symmetric spaces $K_1\backslash T^*G/K_2$ where $G$ is a Lie group and $K_1,K_2\subset G$ are Lie subgroups\footnote{When the nature of the quotient space is not specified in this paper, it is always
assumed to be a GIT quotient.} and prove the superintegrability of such systems under natural assumptions about the structure of orbits of $K_1\times K_2$. In particular we assume that the action of $K_1\times K_2$ by left and right translations on $G$ is free. 
The phase space of such a system 
is a symplectic leaf of $K_1\backslash T^*G/K_2$\footnote{In two examples that we consider the quotient space $K_1\backslash T^*G/K_2$ is singular (a stratified symplectic space \cite{SL}), but it has a Zariski open subset which is a
smooth algebraic orbifold.  A symplectic leave of such space has a similar structure: it
has a  Zariski open subset which is a smooth orbifold.}  and the Hamiltonians are pull-backs of $Ad^*$-invariant functions 
on $\g$ with respect to the natural projection $T^*G\to \g^*$. Here we first trivialize $T^*G\simeq \g^*\times G$ by right translations and then project to the first component. 

Several interesting particular cases of such systems are known. In  \cite{FP1}  the authors studied the case of two-sided spin Calogero-Moser model when $K=K_1=K_2$ is the fixed point subgroup
of the Cartan involution\footnote{The restriction of the Chevalley involution of a simple complex Lie group to its split real split  form
is the Cartan involution} of simple noncompact real Lie group $G$. They described commuting integrals, the projection method
for solving equations of motion and computed Hamiltonians in some examples, including 3-parameter Calogero-Moser type
model related to the $BC_n$ root lattice. These results were generalized in \cite{Fe1} to the case when $K_2$ is a subgroup
of fixed points on any involution of $G$. The relation of these models to the geometry of Higgs bundles was explored in \cite{KLOZ1},\cite{KLOZ2}.

Symplectic leaves $S(\cO_1,\cO_2)$ of  $K_1\backslash T^*G/K_2$
are parametrized by a pair of coadjoint
orbits $\cO_1\subset \kf_1^*$,  $\cO_2\subset \kf_2^*$ for Lie group $K_1$ and $K_2$ respectively. When one of them is trivial such systems are exactly the ones from \cite{Fe}.

An important particular case is when $G$ is a complex simple Lie group and  $K_1=K_2=K=G^\theta$ is the subgroup of fixed points of the Chevalley automorphism $\theta$. We show that in this case the symplectic leaf $S(\cO_1,\cO_2)$ has a Zariski open subset $S_{reg}(\cO_1. \cO_2) \subset S(\cO_1,\cO_2)$
\[
S_{reg}(\cO_1. \cO_2)\simeq (T^*H_{reg}\times \cO_1\times \cO_2)/N_K(H_{reg})
\]
Here $H$ is the Cartan subgroup of $G$, $H_{reg}\subset H$ is the set of regular elements in the Cartan subgroup, i.e.
elements with trivial stabilizer of the natural action of the Weyl group $W$ on $H$. The subgroup $N_K(H_{reg})\subset K$ is the normalizer of $H$ in $K\subset G$\footnote{In the split case $N_K(H_{reg})$ has a normal subgroup $Z_K(H)$ which is the centralizer of $H_{reg}$ in $K\subset G$. The quotient group $N_K(H_{reg})/Z_K(H)$ is naturally isomorphic to the
Weyl group $W$ of $G$.} . 

The pullback of $G$-invariant functions on $\g^*$ gives a Poisson commutative subalgebra of rank $r$. This is the subalgebra of Hamiltonians. 
The Hamiltonian corresponding to the quadratic Casimir is
\[
H_{sCM}=\frac{1}{2}\tr(x^2) = \frac{1}{2}x_0^2 + \sum_{\alpha>0} \frac{(h_\alpha \mu_\alpha + \mu_\alpha') (h_{-\alpha} \mu_\alpha + \mu_\alpha')}{(h_\alpha - h_{-\alpha})^2}
\]
Here $\mu_\alpha, \mu_\alpha'$ are coordinates on $\cO_1, \cO_2$ corresponding
to positive roots, $(p,h)\in T^*H_{reg}$.

Another interesting particular case corresponds to the symmetric pair 
$G\subset G\times G$
where $G$ is embedded diagonally in $G\times G$. In this case spin Calogero-Moser systems are determined by the choice of a pair of coadjont orbits $\cO,\cO'$ in $\g^*$. The phase space of such a system is the symplectic leaf
$S(\cO,\cO')$ of $G\backslash T^*(G\times G)/G$ which is the result of appropriate Hamiltonian reduction. The subalgebra of Poisson commutative Hamiltonians
in this case has rank $2r$ where $r=rank(\g)$. 

The symplectic leaf $S(\cO,\cO')$ has a Zariski open subset
\[
S_{reg}(\cO,\cO')\simeq (T^*H_{reg}\times (\cO\times \cO')\sslash H)/W\subset S(\cO,\cO')
\]
where $(\cO\times \cO')\sslash H)$ is the Hamiltonian reduction of the Cartesian product
of two orbits with respect to the diagonal action of the Cartan subgroup. We compute
Hamiltonians and give a description of the model in terms of local Darboux coordinates, 
when $G=SL_n$ 
and both orbits have rank one. Note that these systems can be regarded as linearization
of integrable systems on moduli spaces of flat connections of a torus with two punctures \cite{AR},
see also \cite{ChF} and references therein.

It would be interesting to describe all triples $(G,K_1,K_2)$ which produce meaningful integrable systems.  

\subsection*{4} This paper is organized as follows. In the first section, we construct 
spin Calogero-Moser type systems on symplectic leaves of $K_1\backslash T^*G/K_2$ and Poisson projections which prove their superintegrability. In the second section, we analyze  the example of symmetric pairs $K_1=K_2=G^\theta$ where $G^\theta$ is the subgroup of
fixed points of the Chevalley automorphism $\theta$ of $G$. In the third section, 
we consider an example of the symmetric pair $G\in G\times G$
where $G$ is a simple Lie group embedded in $G\times G$ diagonally. In this case $K_1=K_2=G$. In section 4,
we describe in detail the two--sided spin Calogero-Moser system corresponding to two rank 1 orbits. 
In the Appendices,  we recall some basic and useful facts on action of
Lie groups on their contangent bundles and Poisson brackets between matrix element functions. In the Conclusion, we outline some further problems.

\subsection*{5}{\bf Acknowedgements}  This work started in collaboration with J. Stokman. The author is grateful  to J. Stokman for numerous illuminating 
discussions and important remarks.
The author also thank S. Artamonov and V. Serganova for fruitful discussions and L. Feher for important remarks. This work was partly supported by the grant NSF DMS-1601947 and by the grant RFBR No. 18-01-00916. 
The paper was finished when the author was visiting ETH-ITS, Zurich (Institute for Theoretical Studies). He is grateful to the Institute for the hospitality.

\section{Two--sided Spin Calogero-Moser systems}
Let $G$ be a Lie group. Its actions on itself by left and right translations being lifted to actions on $T^*G$  are Hamiltonian, assuming $T^*G$ is equipped with the standard 
symplectic form. Thus, such actions of any Lie subgroup are also Hamiltonian. 
Let $K_1, K_2\subset G$ be Lie subgroups. The action of $K_1$ by left translations and of $K_2$ by right translations are also Hamiltonian. Thus, the quotient space $K_1\backslash T^*G/K_2$ (its smooth part) is a Poisson manifold.

The spin Calogero-Moser type systems, which we will construct here,
will be called two--sided because their phase spaces are symplectic leaves
of  (the smooth part of) the two--sided quotient $K_1\backslash T^*G/K_2$. The construction is general, and it works for any Lie group $G$ and a pair of Lie subgroups, provided that the action of $K_1\times K_2$ by left-right translations on some varieties (see below) is free.

\subsection{The phase space}
In what follows, we will use some basic facts about actions of a Lie group $G$ on its
cotangent bundle $T^*G$ which are summarized in the Appendix \ref{cot}.

From now on we trivialize $T^*G \simeq \g^* \times G$ by right translations. Then the natural right $G$-action on $T^*G$ becomes the following action on $\g^*\times G$ (see Appendix \ref{cot})
\begin{equation}\label{L}
 h : (x, g) \mapsto (x, gh\inv)
\end{equation}
The natural left $G$-action becomes:
\begin{equation}\label{R}
  h : (x, g) \mapsto (Ad^*_hx, hg)
\end{equation}
These actions are Hamiltonian with moment maps $\mu^{L,R}:\g^*\times G\to \g^*$:
\[
  \mu_G^R (x, g) = -Ad^*_{g\inv} x , \quad \mu_G^L(x, g) = x
\]

Let $K$ be any Lie subgroup of $G$. Since both the left and the right actions of $G$ on $T^*G$ are Hamiltonian,
so are the left and the right actions of the subgroup $K\subset G$. The corresponding moment maps $\mu_K^{L,R}: \g^*\times G\to \kf^*$, are $\mu_K^{L,R}(x)=\pi\circ\mu_G^{L,R}$,
where $\pi: \g^*\to \kf^*$ is the natural projection dual to the embedding $\kf\subset \g$. 
Therefore, if $K_1,K_2\subset G$ are two Lie subgroups, the double coset $K_1\backslash T^*G/K_2$ is a Poisson variety. 

Denote by $\mu_{12}: T^*G\to \kf^*_1\oplus \kf^*_2$ the moment map
\[
\mu_{12}(x,g)=(\pi_1(x), -\pi_2(Ad^*_{g^{-1}}(x)))
\]
and 
\[
\mu_{12}^{-1}(\cO_1,\cO_2)=(\mu_{K_1}^L)\inv (\cO_1 )\cap (\mu_{K_2}^R)\inv ( \cO_2)
\]
This is a co-isotropic subvariety in $T^*G$.

Symplectic leaves of  $K_1 \backslash T^*G / K_2$ are given by the Hamiltonian reduction as
\begin{align*}
  \cS(\cO_1, \cO_2) &= K_1\backslash \mu_{12}^{-1}(\cO_1,\cO_2) / K_2 \\
  &= K_1 \backslash \{(x, g) \mid \pi_1(x) \in \cO_1, -\pi_2(Ad^*_{g^{-1}}(x)) \in \cO_2\} / K_2
\end{align*}
where $\pi_i : \g^* \to \kf_i^*$ are projections dual to the inclusions $\kf_i\subset \g$. The symplectic manifold $\cS(\cO_1, \cO_2)$ is the phase spaces of the spin Calogero-Moser system associated with $K_1,K_2\subset G$ and coadjoint orbits $\cO_1\subset \kf_1^*$ and  $\cO_1\subset \kf_1^*$. 

Let us compute the dimension of $\cS(\cO_1, \cO_2)$ under the  assumption that
$K_1\times K_2$ {\it acts generically free (i.e. acts free on an open dense subset) on $\mu_{12}^{-1}(\cO_1,\cO_2)$}. It is clear that
under this assumption 
\[
dim(\cS(\cO_1, \cO_2)=dim(\mu_{12}^{-1}(\cO_1,\cO_2))-dim(K_1)-dim(K_2)
\]
Taking into account that 
\[
dim(\mu_{12}^{-1}(\cO_1,\cO_2))=dim(T^*G)-(dim(K_1)-dim(\cO_1))-(dim(K_2)-dim(\cO_2)
\]
we have 
\begin{equation}\label{dimS}
dim(\cS(\cO_1,\cO_2))=dim(T^*G)-2dim(K_1)-2dim(K_2)+dim(\cO_1)+dim(\cO_2)
\end{equation}

\subsection{Hamiltonians of two--sided Spin Calogero-Moser systems} \label{sCM-ham}Let $f$ be a function on $\g^*$ which is invariant with respect to the coadjoint action of $G$, i.e. $f(Ad^*_g(x))=f(x)$ for any $x\in \g^*$ and $g\in G$. As it follows from (\ref{L}) and (\ref{R}) its
pull back to $T^*G\simeq \g^*\times G$ is invariant with respect to the left and the right $K$-action on $T^*G$ for any subgroup $K$ in $G$,
and therefore defines a function on $K_1\backslash T^*G/K_2$. We will call the subalgebra of such functions in the algebra of functions on $K_1\backslash T^*G/K_2$ {\it the subalgebra of Hamiltonians}.

The projection $T^*G\simeq \g^*\times G\to \g^*$ is 
Poisson, therefore the subalgebra of Hamiltonians is a Poisson subalgebra. 
From the explicit form of Poisson brackets on $T^*G$ (Appendix \ref{cot}) it is easy to
see that this Poisson subalgebra is Poisson commutative.

\subsection{Superintegrability of the two-sided spin Calogero-Moser system on $\cS(\cO_1, \cO_2)$}
Here we will construct the projections (\ref{SI}) which prove the superintegrability of the two--sided spin Calogero-Moser system.

\subsubsection{The Poisson variety $\cP(\cO_1, \cO_2)$}
First note that the subset 
\[
(\mu^L_G\times \mu^R_G)(\mu_{12}^{-1}(\cO_1,\cO_2))=\{x,-Ad^*_{g^{-1}}(x)|(x,g)\in \mu_{12}^{-1}(\cO_1,\cO_2)\}\subset \kf^*_1\times \kf^*_2
\]
is invariant with respect to the coadjont $K_1\times K_2$-action. Here $\mu_G^{L,R}$ are moment maps for left and right action of $G$ on $T^*G$.

Define the Poisson variety $\cP(\cO_1, \cO_2)$ as
\[
\cP(\cO_1, \cO_2)=(\mu^L_G\times \mu^R_G)(\mu_{12}^{-1}(\cO_1,\cO_2))/K_1\times K_2
\]
We can also represent it as
\begin{equation*}
  \cP(\cO_1, \cO_2) =  \{(x,-y) \mid\pi_1(x) \in \cO_1,  \pi_2(-y) \in \cO_2, Ad_G^*y = Ad_G^*x\} / K _1\times K_2
\end{equation*}

Because the coadjoint action of $K_1\times K_2$ on $\g^*\times \g^*$ (as a subroup of $G$) is Hamiltonian, the quotient space $\cP(\cO_1, \cO_2)$ is Poisson.

Its dimension can be easily computed if we {\it assume that $K_1\times K_2$-action 
on $(\mu^L_G\times \mu^R_G)(\mu_{12}^{-1}(\cO_1,\cO_2)$ is generically free}:
\begin{align*}
 \dim(\cP(\cO_1, \cO_2)) &= 2dim(\g^*) - dim(\g^*/G)-2dim(K_1)-2dim(K_2)+dim(\cO_1)+dim(\cO_2), \\
 &=dim(\cS(\cO_1, \cO_2))-dim(\g^*/G)
\end{align*}

The mapping $ K_1(x,g)K_2\mapsto (Ad^*_{K_1}\times Ad^*_{K_2})(x,-Ad^*_{g^{-1}}x)$ gives the Poisson projection\footnote{Here and in the rest of the paper we will use notation $V\widetilde{\times}_{\g^*/G}W=\{(x,y)\in V\times W| Ad^*_G(x)=-Ad^*_G(y)\}$.}
\[
K_1\backslash T^*G/K_2\to (\g^*\widetilde{\times}_{\g^*/G} g^*)/K_1\times K_2
\]
Its restriction to the symplectic leaf $\cS(\cO_1, \cO_2)$ gives the natural Poisson projection
\[
p_1: \cS(\cO_1, \cO_2)\to \cP(\cO_1, \cO_2)
\]
This mapping is clearly surjective.

\subsubsection{The variety $ \cB(\cO_1, \cO_2)$}
Define the variety $ \cB(\cO_1, \cO_2)$ as the quotient space
\begin{align*}
  \cB(\cO_1, \cO_2) &= \{G\times G \ \ \text{orbits through} \ \ (\mu_G^R \times \mu_G^L)(\cS(\cO_1, \cO_2))\} \\
  &= \{ G\times G \ \ \text{orbits through} \ \ (x,-y)|\pi_1(x)\in \cO_1, \pi_2(y)\in \cO_2, Ad^*_Gx=Ad^*_Gy \} \\
  &= \{(\cO,-\cO)\in \g^*/G\times \g^*/G|\cO_1\subset \pi_1(\cO), \cO_2\subset \pi_2(\cO)\}\\
  &\simeq \{\cO\in \g^*/G|\cO_1\subset \pi_1(\cO), \cO_2\subset \pi_2(\cO)\}\subset \g^*/G
\end{align*}
In the algebraic case, this variety is the $Spec$ variety for the algebra of spin Calogero Moser Hamiltonians (see section \ref{sCM-ham}) restricted to $\cS(\cO_1, \cO_2)$.

The mapping $(Ad^*_{K_1}(x),-Ad^*_{K_2}(y))\mapsto (Ad^*_G(x), -Ad^*_G(y))$ defines the Poisson surjective projection $p_2: \cP(\cO_1, \cO_2)\to \cB(\cO_1, \cO_2)$.
    
The fiber $\cP(\cO_1, \cO_2, \cO) \subset \cP(\cO_1, \cO_2)$  of $p_2$ over $\cO\in \cB(\cO_1, \cO_2)$  has a natural symplectic structure and is a symplectic leaf of $\cP(\cO_1, \cO_2)$. As a symplectic manifold it can be described through the Hamiltonian reduction :
\begin{equation*}
\cP(\cO_1, \cO_2, \cO) = \{(x,-y) \in \g^* \times \g^* \mid x \in \cO,\ y \in -\cO,\ \pi_1(x) \in \cO_1,\ \pi_2(y)  \in -\cO_2\}/K_1\times K_2
\end{equation*}
We have 
\[
\cP(\cO_1, \cO_2, \cO) = \cM_1(\cO,\cO_1)\times \cM_2(-\cO,-\cO_2)
\]
Here both $\cM_1(\cO,\cO_1)$ and  $\cM_2(-\cO,-\cO_2)$ are symplectic manifolds with the symplectic
structure given by the Hamiltonian reduction of $\cO\subset \g^*$ with respect to the coadjoint action of
$K_1$ and $K_2$ respectively\footnote{Note that when $G$ is simple, compact, and when $K_1\subset G$ is simple, the algebra of functions on $\cM_1(\cO,\cO_1)$ has a natural quantization. Consider an irreducible finite dimensional representation $V^\lambda_G$ of $G$ corresponding 
to the orbit $\cO$. As a $K_1$-module it decomposes into irreducible $K_1$-modules:
\[
V^\lambda_G\simeq \oplus_\nu W^\lambda_\nu\otimes V^\nu_{K_1}
\] 
The endomorphism algebra of $W^\lambda_\nu$ is the quantization of functions on $\cM_1(\cO,\cO_1)$,
such that $\lambda$ corresponds to $\cO$ and $\nu$ corresponds to $\cO_1$. Similar quantization has
the algebra of functions on $\cM_2(\cO,\cO_2)$. The compactness and finite dimensionality of $V_G^\lambda$ assumptions are not  important and can be relaxed.}. 

We will say that our Hamiltonian system has {\it full rank} if $dim(\cB(\cO_1,\cO_2)=dim(\g/G)$. This is the case for
the two examples we will consider in this paper.

\subsubsection{Superintegrability}

Thus, we have Poisson projections 
\[
\cS(\cO_1, \cO_2)\stackrel{p_1}\rightarrow \cP(\cO_1, \cO_2)\stackrel{p_2}\rightarrow \cB(\cO_1, \cO_2)\subset \g^*/G
\]
With all our assumptions, the dimensions match and we have a superintegrable system on each symplectic leaf
of $K_1\backslash T^*G/K_2$.

\subsection{Exact solution to equations of motions and factorization}
Let $f\in C(\g^*)^G$ be an $Ad^*_G$-invariant function on $\g^*$.
From the explicit form of the symplectic structure $T^*G$ one can easily derive
that the Hamiltonian flow line generated by the pull back of $f$ to $T^*G\simeq \g^*\times G$
passing through $(x,g)\in T^*G$ at $t=0$ is:
\begin{equation}\label{A}
(x(t), g(t))=(x, \exp(t\nabla f(x))g)
\end{equation}
Assume that $K_1(x,g)K_2\in \cS(\cO_1, \cO_2)$, where $K_1\times K_2$ action is the left-right action:
\[
(k_1,k_2)(x,g)=(Ad^*_{k_1}(x), k_1gk_2) .
\] 
Then the flow line (\ref{A}) projects to  the flow line of the spin Calogero Moser system \\$K_1(x, \exp(t\nabla f(x))g)K_2$,
which stays in $\cS(\cO_1, \cO_2)$ for all $t$. Thus, solutions to corresponding (quite complicated) nonlinear ordinary 
differential equations are obtained by computing coset representatives of (\ref{A}), which is a manageable  algebraic problem.

The Hamiltonian flow of a two-sided spin Calogero-Moser system is complete with the appropriate positivity (or compactness) assumption. In general solutions may develop singularities at finite time. This phenomenon was studied for Toda model in \cite{FT} and in \cite{Kod}.  

\section{Two--sided Calogero-Moser systems for symmetric pairs of Cartan type } \label{sCMK} 

Here we will construct superintegrable systems of spin Calogero-Moser type corresponding to symmetric pairs $K\subset G$, where  $G$ is a simple complex Lie group and $K=G^\theta \subset G$ is the subgroup of fixed points of the Chevalley automorphism $\theta: G\to G$.

Recall that if we choose a Borel subroup $B\subset G$ and if $H\subset B\subset G$ is the
Cartan subgroup of $G$, the Chevalley automorphism (corresponding to this choice) acts on the Lie algebra
$\g=Lie(G)$ as $\theta(h)=-h$, $\theta(e_\alpha)=-e_{-\alpha}$ where $h$ is an element of the Cartan subgroup $\hf\subset \g$ and $\{e_\alpha\}$ are Chevalley generators corresponding to roots of $\g$ . We will denote roots by $\Delta$ and positive roots by
$\Delta_+$.  Note that elements $e_\alpha-e_{-\alpha}$ form a basis in
the Lie subalgebra $\kf\subset \g$.

When $G$ is a complex algebraic group there are subtleties related to the fact that 
the double coset space $K\backslash G/K$ is singular outside of the Zariski open
subset $KHK\subset G$. We will not be going into these subtleties here since in most important
cases when $G$ is the split real form $G_\RR$ or when $G$ is the maximal 
compact subroup of $G_\CC$ there exists global spherical decompositions
$G_\RR=KAK$ and $G=KHK$.

\subsection{The regular part of $\cS(\cO_1, \cO_2)$}
Let $\kf\subset \g$ be the Lie subalgebra of fixed points of $\theta$. After the identification of $\g^*$ with $\g$ by the Killing form, the projection $\g^*\to \kf^*$ acts as
\[
  \pi(x_0 + \sum_{\alpha\in \Delta} x_\alpha e_\alpha) = \sum_{\alpha\in \Delta_+} (x_\alpha-x_{-\alpha}) (e_\alpha - e_{-\alpha})  .
\]
where $x_0\in \hf$ is the Cartan component of $x$.

We will say that $g\in G$ is {\it regular} (in the context of the symmetric pair $K\subset G$ that we consider here) if 
it can be written as $g=k_1 g k_2$ where $k_1,k_2\in K$ and 
$h\in H_{reg}$\footnote{Recall, that $h\in H$ is {\it regular} if 
the stabilizer of $h$ is $W$ is trivial, $H_{reg}\subset H$ is the subset of regular elements.
If $G_\RR$ is the split real form 
of $G$ and $K\subset G_\RR$ is the maximal compact Lie subgroup in $G$ we have the polar decomposition
$G_\RR=KAK$ where $A$ is the positive part of the Cartan subroup in $G_\RR$. When $G$ is compact real form of $G_\CC$ we also have a global decomposition $G=KTK$,
where $T$ is the Cartan subroup of $G$.}. We will denote $G_{reg}\subset G$ the subset of regular elements.
Thus, we have a Zariski open subset $K\backslash T^*G_{reg}/K\subset K\backslash T^*G/K$ which is isomorphic to $(\g^*\times H_{reg})/N_K(H_{reg})$ where $N_K(H_{reg})\subset K$ is the normalizer of
the regular part Cartan subroup $H_{reg}\subset G$ in $K\subset G$. The action of $N_K(H_{reg})$ is diagonal with the coajoint action on the first factor and with the natural action of the Weyl group on the second factor\footnote{ The group $N_K(H)$ has a normal subgroup $Z_K(H)$ which is the centralizer of $H$ in $K$, and there is a natural isomorphism $N_K(H)/Z_K(H)\simeq W$, where $W$ is the Weyl group of $G$.}.

In the Appendix \ref{KG} we described the regular part $K\backslash G_{reg}/K\subset K\backslash T^*G/K\simeq (\g^*\times H_{reg})/N(H_{reg})$ as a Poisson variety which is isomorphic to $(\kf^*\times \kf^*\times T^*H_{reg})/N_K(H_{reg})$. In particular, we obtained
formulae expressing root coordinates of $x\in \g^*$ in terms of $(\mu, \mu')\in \kf^*\times \kf^*\subset \kf^*\times \kf^*\times H_{reg}$:
\begin{equation}\label{xfncn}
x_\alpha = \frac{\mu_\alpha' + h_{\alpha} \mu_\alpha}{h_{\alpha} - h_{-\alpha}}, \ \ x_\alpha = \frac{\mu_\alpha' + h_{-\alpha} \mu_\alpha}{h_{\alpha} - h_{-\alpha}},
\end{equation}
The Cartan component $x_0$ in this identification becomes a cotangent vector in $T^*H_{reg}$.

Specializing this isomorphism to the symplectic leaf $S(\cO_1, \cO_2)$ where $\cO_1, \cO_2\subset \kf^*$ are coadjoint orbits, we obtain:

\begin{equation}\label{biso}
S_{reg}(\cO_1,\cO_2)\simeq ( \cO_1 \times \cO_2\times T^*H_{reg})/N_K(H_{reg}) \subset \cS(\cO_1, \cO_2)
\end{equation}  
As it follows from this formula, the dimension of $\cS(\cO_1, \cO_2)$  is $2r+dim(\cO_1)+dim(\cO_2)$, agrees naturally with the general formula (\ref{dimS}). 

The symplectic form on $\cS_{reg}(\cO_1, \cO_2)$ is the pull-back of the symplectic form
$\omega_1+\omega_2+\omega_3$ where $\omega_1$ is the canonical symplectic form on $T^*H_{reg}$,
$\omega_2$ and $\omega_3$ are Kirillov-Kostant symplectic forms on  $\cO_1$ and $ \cO_2$ respectively.

\subsection{The Hamiltonians} Now let us describe Hamiltonians of the spin Calogero-Moser system in terms of these coordinates.
Clearly there exists $r=rank(G)$ independent Hamiltonians. They can be chosen as $\tr_V(\pi(x)^k)$ where $k=2.3,\dots$
where $x=x_0+\sum_\alpha x_\alpha e_\alpha\in \g\simeq\g^*$ evaluated in a finite dimensional representation $(\pi, V)$ of $G$
and $x_\alpha$ are given by (\ref{xfncn}).

The spin Calogero--Moser Hamiltonian on $\cS(\cO_1, \cO_2)$ corresponds to the second Casimir and is
\begin{equation}\label{HCM}
 H_{sCM}=\frac{1}{2}\tr_{Ad}(x^2) = \frac{1}{2}x_0^2 + \sum_{\alpha>0} \frac{(h_\alpha \mu_\alpha + \mu_\alpha') (h_{-\alpha} \mu_\alpha + \mu_\alpha')}{(h_\alpha - h_{-\alpha})^2}
\end{equation}
where the trace is taken over the adjoint representation.

\begin{remark} 
Note that in the description  of birational coordinates $x_0, h_\alpha, \mu, \mu'$ we used the 
spherical decomposition $KHK\subset G$ which gives only a Zariski open subvariety in complex algebraic case. 

In the real split case, the polar decomposition $G_\RR=KAK$ is global and we can use real coordinates 
$h_\alpha=\exp(\frac{1}{2}q_\alpha)$

\[
 H_{sCM}=\frac{1}{2}\tr(x^2) = \frac{1}{2}p^2 + \sum_{\alpha>0} \frac{\mu_\alpha^2+ 2ch(\frac{1}{2}q_\alpha)\mu_\alpha'\mu_\alpha + \mu_\alpha'^2}{4sh^2(\frac{1}{2}q_\alpha)}
\]
In the compact case it is also global and 

\[
 H_{sCM}=\frac{1}{2}\tr(x^2) = -\frac{1}{2}p^2 - \sum_{\alpha>0} \frac{\mu_\alpha^2+ 2cos(\frac{1}{2}q_\alpha)\mu_\alpha'\mu_\alpha + \mu_\alpha'^2}{4sin^2(\frac{1}{2}q_\alpha)}
\]

\end{remark}

\begin{remark}
The linear operator 
\[
x=x_0+\sum_\alpha x_\alpha e_\alpha
\]
whose matrix elements are functions on $\cS_{reg}(\cO_1, \cO_2)$ is the Lax operator
of the spin Calogero Model.
\end{remark}

\subsection{Solutions to equations of motion}

If the spherical factorization is global, the flow line can be described as
follows. First write 
\[
\exp(t\nabla f(x))g=k_1(t)a(t)k_2(t)
\]
Now note that each class $K(x,g)K$ has a representative $(Ad_{k_1^{-1}}^*x, a)$ where $a\in A$,
$k_1,k_2\in K$ and $g=k_1ak_2$. Such representative of the class $K(x(t),g(t))K$ is $(Ad^*{k_1(t)^{-1}}(x),a(t)$.
In therm of coordinates (\ref{biso}) we have $a(t)_\alpha=\exp(\frac{1}{2}q_\alpha(t))$ while $p(t), \mu(t), \mu(t)'$
are defined as
\[
Ad_{k_1(t)^{-1}}(x)=p(t)+\sum_{\alpha\in \Delta_+}\frac{\mu'_\alpha(t)+a_{-\alpha}(t)\mu_\alpha(t)}{a_{\alpha}(t)-a_{-\alpha}(t)}e_\alpha+\sum_{\alpha\in \Delta_+}\frac{\mu'_{-\alpha}(t)+a_{\alpha}(t)\mu_{-\alpha}(t)}{a_{-\alpha}(t)-a_{\alpha}(t)}e_{-\alpha}
\]
Here the initial values of coordinates $p,\mu,\mu'$ are determined by $x$:
\[
x=p+\sum_{\alpha\in \Delta_+}\frac{\mu'_\alpha+a_{-\alpha}\mu_\alpha}{a_{\alpha}-a_{-\alpha}}e_\alpha+\sum_{\alpha\in \Delta_+}\frac{\mu'_{-\alpha}+a_{\alpha}\mu_{-\alpha}}{a_{-\alpha}-a_{\alpha}}e_{-\alpha}
\]

\subsection{One--sided spin Calogero-Moser systems}
If one of the coadjoint orbits, either $\cO_1$ or $\cO_2$, is trivial, we will say that the
corresponding spin Calogero--Moser model is {\it one--sided}. In this case, up to left--right equivalence
the phase space of such a system is a symplectic  leaf of $K_1\backslash T^*(G/K_2)$ which can be described as
\[
\cS(\cO)=\cS(\cO,\{0\})=K_1\backslash \mu_{K}^{-1}(\cO)
\]
where $\mu_{K}: T^*(G/K_2)\to \kf_1^*$ is the moment map for the left $K_1$ action on $T^*(G/K_2)$.

When $K_1=K_2=K$ is the subgroup of fixed points of the Chevalley involution, the regular part of $S(\cO)$, just as for the two--sided model, is s Zariski open subset 
\[
S_{reg}(\cO)\simeq (T^*H_{reg}\times \cO^K)/N_K(H) \subset \cS(\cO)
\]
This embedding is a birational equivalence of symplectic varieties.

The Hamiltonian of the one--sided spin Calogero-Moser system in this case is the specialization of (\ref{HCM}):
\[
 H_{sCM}=\frac{1}{2}\tr(x^2) = \frac{1}{2}x_0^2 + \sum_\alpha \frac{\mu_\alpha^2}{(h_\alpha - h_{-\alpha})^2}
\]
The real form of this model was discussed in \cite{Fe}.

\section{Spin Calogero-Moser systems on $G \times G$}

Here we will focus on the spin Calogero-Moser type systems corresponding  to the symmetric space $G \times G/G$, where $G\subset G\times G$ is the subgroup of fixed points of the permutation $(x,y)\mapsto (y,x)$, i.e.
the Lie group $G$ is embedded diagonally in $G\times G$.  These systemes are examples of the general construction from the first section where $\theta: G \times G\to G\times G$ is the permutation $\theta(g_1, g_2) = (g_2, g_1)$. 

We trivialize $T^*(G \times G)$ by right translations to $\g^* \times \g^*\times G \times G = \{(x, y, g, h)\}$, where $(x, g)$ and $(y, h)$ are elements of two copies of $T^*G\simeq \g^*\times G$. From now on we assume that $G$ is simple.

\subsection{The phase space}
The phase space $\cS(\cO,\cO')$ of such a system is parameterized by a pair of coadjoint orbits $\cO, \cO'\subset \g^*$ and is a symplectic leaf of the Poisson variety $G\backslash T^*(G \times G)/G$ where $G$ is diagonally embedded
into $G\times G$. It is given by the Hamiltonian reduction
\begin{align}\label{sl-GxG}
  \cS(\cO,\cO') &= G\backslash (\mu^L_G)^{-1}(\cO)\cap (\mu^R_G)^{-1}(\cO')/G\nonumber \\ 
  &=G\backslash \{(x,y,g,h)\in \g^* \times \g^*\times G \times G|x+y\in \cO, -Ad^*_{g^{-1}}(x)-Ad^*_{h^{-1}}(y)\in \cO'\} \\
  &= G\backslash \{(x, y, a) \mid x + y \in \cO, x + Ad^*_{a} y  \in -\cO'\}\nonumber
\end{align}
Here $\mu_G^{L,R}$ are moment maps for the left and right diagonal action of $G$
on $T^*(G\times G)$, see Appendix \ref{GG}.

\begin{remark}
The symplectic leaf  $\cS(\cO,\cO')$ can also be thought of as a symplectic leaf of the Poisson 
variety $G\backslash \cS(\cO')$
where $\cS(\cO')=(\mu^R_G)^{-1}(\cO')/G$ is a symplectic leaf of the Poisoon variety $T^*(G\times G)/G$.
The Poisson structure on it is determined by the natural symplectic structure on $T^*G$ and by the Hamiltonian action of $G$. Because the left $G$ action on $\cS(\cO')$ is Hamiltonian, the quotient space $G\backslash \cS(\cO')$ is Poisson, and there exists a moment map $\mu: S(\cO')\to \g^*$. Symplectic leaves of $G\backslash \cS(\cO')$ are given by the
Hamiltonian  reduction and $\cS(\cO,\cO')=G\backslash \mu^{-1}(\cO)$. For details see
Appendix \ref{GG}. 
\end{remark}

It is easy to compute the dimension of $\cS(\cO,\cO')$:
\begin{equation}\label{dsoo}
dim(\cS(\cO,\cO'))=dim(\cO)+dim(\cO')
\end{equation}
Note that this naturally agrees with (\ref{dimS}).

\subsection{The Hamiltonians}
After the trivialization of the cotangent bundle $T^*(G\times G)$ by right translations $T^*(G\times G)\simeq \g^*\times \g^*\times G\times G$ we have a natural projection $T^*(G\times G)\to \g^*\times \g^*$. This projection
gives the projection $G\backslash T^*(G\times G)/G\to \g^*/G\times \g^*/G$. The restriction of this projection to
the symplectic leaf $\cS(\cO,\cO')$ gives the projection $\cS(\cO,\cO')\to \cB(\cO,\cO')\subset \g^*/G\times \g^*/G$,
where $\cB(\cO,\cO')$ is simply the image of the symplectic leaf. It will be described more explicitly later.

The algebra of Hamiltonians of the two--sided spin Calogero Moser system for $G\subset G\times G$ is
the pullback of functions on $\g^*/G\times \g^*/G$, restricted to $\cB(\cO,\cO')\subset \g^*/G\times \g^*/G$, to functions on $\cS(\cO,\cO')$.

\subsection{The superintegrability of the two--sided spin Calogero Moser system on $\cS(\cO, \cO')$}

\subsubsection{The Poisson variety $\cP(\cO, \cO')$}
The Lie group $G \times G$ acts on $\cS(\cO)$ from the left and from the right. Let $\mu_{G\times G}^L$ and $\mu_{G\times G}^R$ be corresponding moment maps $\mu_{G\times G}^L,\mu_{G\times G}^R: \cS(\cO, \cO')\to \g^*\times \g^*$. 

Define $\cP(\cO, \cO')$ as the space of $G\times G$ orbits in $\g^*\times \g^*\times \g^*\times \g^*$ passing through $\mu_{G\times G}^L\times \mu_{G\times G}^R( \cS(\cO, \cO'))$ where the first copy of $G$ acts diagonally on the
first two copies on $\g^*$ and the second copy of $G$ acts diagonally on the last two copies of $\g^*$. In other words:

\begin{align*}
  \cP(\cO, \cO') = \{(x, y,u,v)\in  (\g^*)^{\times 4}\mid x+y\in \cO, u+v\in \cO', \\ Ad^*G(u)=-Ad^*_G(x), Ad^*G(v)=-Ad^*_G(y)\}/G\times G 
\end{align*}
where $(g,h)(x,y,u,v)=(Ad^*_g(x),  Ad^*_g(y),  Ad^*_h(u), Ad^*_h(v))$.
It is easy to compute its dimension for generic orbits $\cO,\cO'$:
\begin{equation*}
  \dim(\cP(\cO, \cO'))=dim(\cS(\cO, \cO'))-2dim(\g^*/G)
\end{equation*}

The mapping 
\[
p_1: G(x,y,g,h)G\mapsto (G\times G)(x,y, -Ad^*_{g^{-1}}(x), -Ad^*_{h^{-1}}(y))
\]
is the Poisson projection $p_1: \cS(\cO, \cO')\to \cP(\cO, \cO')$.

\subsubsection{The variety $\cB(\cO, \cO')$} In the algebraic case the variety $\cB(\cO, \cO')$ is the {\it Spec} variety of
the Poisson center of the Poisson algebra of functions on $\cP(\cO, \cO')$. It can be described explicitly as follows:
\[
\cB(\cO, \cO')=\{\cO_1,\cO_2\in \g^*/G| \cM(\cO_1,\cO_2|\cO)\neq \emptyset,  \cM(\cO_1,\cO_2|-\cO')\neq \emptyset\}
\]
Here $\cM(\cO_1,\cO_2|\cO)=\{x\in \cO_1, y\in \cO_2| x+y\in \cO\}$. We have $dim(\cB(\cO, \cO'))=2 dim(\g^*/G)=2r$ where $r$ is the rank of $\g$.

\subsubsection{Superintegrability}

Consider the  mapping $(\g^*)^{\times 4}/Ad^*_{G\times G}\to (\g^*/G)^{\times 4}$ acting as
\begin{align*}
\Ad_{G\times G}^* (x, y, u, v) \mapsto (Ad_G^*(x), Ad_G^*(y), Ad_G^*(u), Ad_G^*(v))=\\ (Ad_G^*(x), Ad_G^*(y),-Ad_G^*(x), -Ad_G^*(y).
\end{align*} 
The image is determined by $\cO_1= Ad_G^*(x)$ and $\cO_2=Ad_G^*(y)$ and therefore defines an element  $\g^*/G\times \g^*/G$. This gives a surjective mapping $(\g^*)^{\times 4}/Ad^*_{G\times G}\to \g^*/G\times \g^*/G$.
Restricting this map to $\cP(\cO, \cO')$ we obtain the surjective projection
\begin{align*}
p_2:  \cP(\cO, \cO') \to B(\cO, \cO') 
\end{align*}
which is clearly Poisson if we equip the base $B(\cO, \cO')$ with the trivial Poisson structure.

It is easy to see that the fiber of this projection over $(\cO_1,\cO_2)$ is 
\begin{align*}
  \cS(\cO_1,\cO_2, \cO', \cO'') = \{(x, y, u,v) \mid x + y \in \cO,\ u+ v \in \cO',\ x, -u\in \cO_1, y, -v\in \cO_2\} / \Ad_{G\times G}^*
\end{align*}
This is a symplectic leaf of $\cP(\cO, \cO')$. Note that
\[
\cS(\cO_1,\cO_2, \cO', \cO'')=\cM(\cO,\cO_1,\cO_2)\times \cM(\cO',-\cO_1,-\cO_2)
\]
where $\cM(\cO,\cO_1,\cO_2)$ is the Hamiltonian reduction of the symplectic manifold
$\cO\times \cO_1\times \cO_2$ with respect to the diagonal coadjoint action of $G$\footnote{Note that the algebra of functions on $\cM(\cO,\cO_1,\cO_2)$ is the classical analog of  the algebra of endomorphisms of the multiplicity space $W^{\lambda_1,\lambda_2}_\lambda$ in the decomposition of two irreducible $G$-modules:
\[
V_G^{\lambda_1}\otimes  V_G^{\lambda_2}\simeq \oplus_{\lambda} W^{\lambda_1,\lambda_2}_\lambda\otimes   V_G^{\lambda}
\]
where $\lambda_1$ corresponds to $\cO_1$, $\lambda_2$  to $\cO_2$
and $\lambda$ to $\cO$ in the usual way of how coadjoint orbits correspond to irreducible representations.}
:
\[
\cM(\cO,\cO_1,\cO_2)=\{(x,y,z)\in \cO\times \cO_1\times \cO_2|x-y-z=0\}/G
\]

Thus, we have two Poisson projections
\[
  \cS(\cO, \cO') \stackrel{p_1}\rightarrow \cP(\cO, \cO') \stackrel{p_2}\rightarrow \cB(\cO, \cO')
\]
with the right balance of dimensions:
\[
  \dim(\cP( \cO, \cO')) = \dim \cS(\cO, \cO') - \dim \cB(\cO, \cO')
\]
That is, we have a superintegrable system on  $\cS(\cO, \cO')$ with Hamiltonians described earlier.

The exact solution of equations of motion via factorization is the specialization to the case when $K_1=K_2=G\subset G\times G$ of the general scheme explained in the first section. The factorization problem is based on the following structure of double cosets $G\backslash (G\times G)_{reg}/G\simeq G_{reg}/Ad_G\supset H/W$. Recall that when $G$ is compact simple $G/Ad_G\simeq H/W$.

\subsection{The regular part of $\cS(\cO,\cO')$}
Here we will describe the subvariety of regular points of $\cS(\cO,\cO')$.
For a regular point $a=ghg^{-1} \in G$\footnote{Recall that $a\in G$ is diagonalizable if $a=ghg^{-1}$ for
some $g\in G$ and $h\in H$. It is {\it regular} if it is diagonalizable and if the Weyl group acts on $h$ without stabilizers.}, where $h\in H_{reg}$. In this case we have a natural identification
\[
G (x, y, a) = N(H_{reg})(x, y, h).
\]
where $N(H_{reg})\subset G$ is the normalizer of $H_{reg}\subset H\subset G$.

As it is shown in Appendix \ref{GG}, there are Poisson isomorphisms
\[
G\backslash T^*(G\times G)/G\simeq (\g^*\times \g^*\times G)/G
\]
and 
\begin{equation}\label{iso-reg}
(\g^*\times \g^*\times G_{reg})/G\simeq (\g^*\times \g^*)\sslash H\times T^*H_{reg}
\end{equation}
where $(\g^*\times \g^*)\sslash H$ is the Hamiltonian reduction with respect to the coadjoint diagonal action of $H$.

Restricting the isomorphism (\ref{iso-reg}) to the regular part of the symplectic leaf $S(\cO, \cO')$ we obtain an isomorphism
\[
\cS(\cO, \cO')_{reg} \simeq \bigl( (\cO \times \cO') \sslash H \times T^* H_{reg} \bigr) / W  
\]

Let $\mu = \mu_0 + \sum_\alpha \mu_\alpha e_\alpha$ and $\mu' = \mu_0' + \sum_\alpha \mu_\alpha' e_\alpha$, $\mu_0, \mu'_0 \in \hh^*$,
be the root decomposition  of elements in $\cO,\cO'\subset \kf^*$.
In the Appendix \ref{GG} we derived formulae expressing components of $x,y\in \g^*\times \g^*\subset \g^*\times \g^*\times G_{reg}$ in terms of $\mu,\mu'$ and $(p,h)\in T^*H_{reg}$.
Specializing these formulae to $\cS(\cO, \cO')_{reg}$ we have:
\[
 x_0=\mu_0 + p, \ \ y_0=- p, \qquad y_\alpha = \frac{\mu_\alpha + \mu'_\alpha}{1-h_\alpha}, \qquad x_\alpha = \frac{\mu_\alpha h_\alpha + \mu'_\alpha}{h_\alpha - 1}.
\]
Here we should take into account that $(\mu,\mu')\in \mu_0^{-1}(0)\subset \cO\times \cO'$, where $\mu_0: \cO\times \cO'\to \hf^*$ is the moment map for the diagonal coadjoint action of the Cartan subgroup $H\subset G$.
This results in the condition $\mu_0 = - \mu_0'$.

From this description of
$\cS(\cO,\cO')$ we have 
\begin{equation*}
\dim(\cS_{reg}(\cO,\cO')) = dim((\cO\times \cO')\sslash H)-2r=dim(\cO)+dim(\cO')
\end{equation*}
which agrees with the formula (\ref{dsoo}) for the dimension of $\cS(\cO, \cO')$.

It is easy to compute the restriction of quadratic Casimir Hamiltonians to $\cS_{reg}(\cO,\cO')$:
\begin{equation}\label{hx}
H^{(x)}=\frac{1}{2}(x,x)=\frac{1}{2}(\mu_0+p,\mu_0+p)+\sum_{\alpha\in \Delta_+} \frac{(\mu_\alpha h_\alpha+\mu_\alpha')(\mu_{-\alpha}h_{-\alpha}+\mu_{-\alpha}')}{(h_{\alpha}-1)(h_{-\alpha}-1))}
\end{equation}
\begin{equation}\label{hy}
H^{(y)}=\frac{1}{2}(y,y)=\frac{1}{2}(p,p)+\sum_{\alpha\in \Delta_+} \frac{(\mu_\alpha +\mu_\alpha')(\mu_{-\alpha}+\mu_{-\alpha}')}{(h_{\alpha}-1)(h_{-\alpha}-1))}
\end{equation}

The variables $x_0$ and $y_0$ are as above, $\mu$ and $\mu'$ satisfy constraints 
\[
c_k(\mu)=c_k, \ \ c_k(-\mu')=c_k', \ \ \mu_0=-\mu'
\]
Here $c_k(x)$ are generators of the algebra of $Ad^*_G$-invariant polynomial functions on $\g^*$,
$c_k$ are values of these functions on $\cO$ and $c_k'$ are their values on $\cO'$.

When $\cO'=\{0\}$ (or $\cO=\{0\}$) the Hamiltonian system described above 
degenerates into the "usual" spin Calogero-Model on orbit $\cO$ (or $\cO'$).
In this case $\mu_0=\mu_0'=0$, $H^{(x)}=H^{(y)}$ and the superintegrability of
the two--sided spin Calogero-Moser system becomes the superintegrability of the spin Calogero-Moser system described in \cite{R2}.

\section{Two--sided spin Calogero-Moser model for rank one orbits for $SL_n$}

\subsection{Rank one orbits for $SL_n$} Assume that $G=SL_n$ and both coadjoint orbits $\cO,\cO'\subset \g^*=sl_n^*$ have rank one. First let us describe an open dense subset in such an orbit as a symplectic leaf of $T^*\CC^n/\CC^*$. In terms of coordinate functions
this realization is given by the following formulae:
\begin{equation}\label{mureal}
\mu_{ij}=\phi_i\psi_j-\frac{\delta_{ij}}{n}(\phi,\psi), \ \ \mu_{ij}'=\phi_i'\psi_j'-\frac{\delta_{ij}}{n}(\phi',\psi')
\end{equation}
where  $(\phi,\psi)=\sum_{i=1}^n\phi_i\psi_j$ and nonzero Poisson brackets between $\phi,\psi,\phi',\psi'$ are
\[
\{\phi_i,\psi_j\}=\delta_{ij}, \ \ \{\phi'_i,\psi'_j\}=\delta_{ij}
\]
Functions $\psi_i\phi_j$ generate the subalgebra of functions on $T^*\CC^n\simeq \CC^n\times \CC^n$ (with $\phi_i$
being functions on cotangent spaces and $\psi_i$ being functions on $\CC^n$) invariant with respect to the 
following action of $\CC^*$: $\phi_i\mapsto \lambda \phi_i$, $\psi_i\mapsto \lambda^{-1} \psi_i$.

Fixing Casimir functions in the rank one case is equivalent to  fixing $N=(\phi,\psi)$.
We will write $\cO_N$ for such orbit. Similarly we will write $\cO_{N'}$ for the orbit $\cO'$ with $N'=(\phi',\psi')$. Note that each orbit has is  $2n-2$-dimensional. This is due to the constraints fixing the orbit and  symmetry $\phi_i\mapsto \lambda \phi_i$, $\psi_i\mapsto \lambda^{-1} \psi_i$, $\phi_i'\mapsto \mu \phi_i'$, $\psi_i'\mapsto \mu^{-1} \psi_i'$. 
In other words, both orbits are described now as the Hamiltonian reduction 
of $T^* \CC^n\sslash \CC^*$.

\subsection{The Hamiltonian reduction $(\cO\times \cO')\sslash H$}
The Cartan subgroup $H\subset SL_n$ acts diagonally on $\cO\times \cO'$.
In terms of variable $\phi, \psi, \phi',\psi'$ this action is
\[
\phi_i\mapsto \mu_i \phi_i, \psi_i\mapsto \mu_i^{-1} \psi_i, \phi_i'\mapsto \mu_i \phi_i', \psi_i'\mapsto \mu_i^{-1} \psi_i'
\]
where $\mu_i\in \CC^*$ and $\mu_1\cdots m_n=1$.

This action is Hamiltonian with the moment map $\mu_0(\mu, mu')=\mu+\mu'$. The Hamiltonian reduction $\mu_0^{-1}(0)/H$ in terms of $\phi, \psi, \phi',\psi'$
means we have to impose the constraint $\mu_{ii}+\mu'_{ii}=0$ and take $H$-invariant functions of
these variables. The constrain implies that
\[
\phi_i\psi_i+\phi'_i\psi'_i=\frac{N+N'}{n}
\]
where $N=\sum_i \phi_i\psi_i$ and $N'=\sum_i \phi'_i\psi'_i$.

The $H$-invariant subalgebra of polynomial functions on $\cO\times \cO'$ consists on 
polynomials which depend only on zero weight combinations of $\mu_{ij}$ and $\mu'_{ij}$.
These combinations can be written in terms of variables
\[
a_i=\phi_i\psi'_i, \ \ b_i=\phi'_i\psi_i, \ \ n_i=\phi_i\psi_i, \ \ n'_i=\phi'_i\psi'_i
\]
For example:
\[
\mu_{ij}\mu_{ji}=n_in_j, \ \  \mu_{ij}\mu'_{ji}=a_ib_j, \ \ \mu'_{ij}\mu'_{ji}=n'_in'_j
\]
The variable $a_i, b_i, n_i, n'_i$ are not independent and they satisfy relations $n_in'_i=a_ib_i$.
Thus, we have the isomorphism of commutative algebras
\[
C^H(\cO\times \cO')\simeq <a_i,b_i,n_i,n'_i| a_i, b_i, n_i, n'_i>/<n_in'_i-a_ib_i>
\]

It is easy to describe Posson brackets between variables $a_i,b_i,n_i,n'_i$. Nonzero Poisson brackets are:
\[
\{a_i, b_i\}=n_i-n'_i, \ \ \{n_i,a_i\}=a_i, \ \ \{n_i,b_i\}=-b_i, \{n'_i,a_i\}=-a_i, \ \ \{n'_i,b_i\}=b_i, 
\]

Denote $H_i=n_i-n_i'$. We can now describe explicitly the ring of functions on $(\cO\times \cO')\sslash H$.

\begin{proposition} The ring of functions on $(\cO\times \cO')\sslash H$ is isomorphic to $\CC^*$-invariant subalgebra of commutative algebra
generated by elements $a_i, b_i, H_i$ with defining relations
\begin{equation}\label{constr}
a_ib_i+\frac{H_i^2}{4}=\frac{(N+N')^2}{4n^2}, \ \ \sum_{i=1}^n H_i=N-N', 
\end{equation}
The action of $\CC^*$ on generators  given by $a_i\mapsto \mu a_i, b_i\mapsto \mu^{-1}b_i, H_i\mapsto H_i$. This algebra has a Poisson structure defined by the following nonzero Poisson brackets
\[
\{a_i, b_i\}=H_i, \ \ \{H_i,a_i\}=2a_i, \ \ \{H_i,b_i\}=-2b_i
\]
\end{proposition}

Thus, we have a birational isomorphism $(\cO\times \cO')\sslash H\simeq (\cO_0^{\times n})_{N-N'}$, where 
$\cO_0$ is the coadjoint orbit of $SL_2$ with the value of the Casimir function determined by $N+N'$ as
in (\ref{constr}), and $(\cO_0^{\times n})_{N-N'}$ is the intersection of $\cO_0^{\times n}$ and
the level set $\sum_{i=1}^nH_i=N-N'$.

Note that we can write $n_i$ and $n'_i$ in terms of $H_i$ and $N, N'$ as
\begin{equation}\label{nN}
n_i=\frac{1}{2}(H_i+\frac{N+N'}{n}), \ \ n'_i=\frac{1}{2}(-H_i+\frac{N+N'}{n})
\end{equation}

As a consequence, for rank one orbits we can write Hamiltonians corresponding to quadratic Casimirs as
\[
H^{(x)}=\frac{1}{2}\sum_{i=1}^n(p_i+H_i+\frac{N-N'}{n})^2+\sum_{i<j} \frac{(n_in_j+n'_in'_j+a_ib_jh_ih_j^{-1}+b_ia_jh_jh_i^{-1})}{(h_i-h_j)(h_i^{-1}-h_j^{-1})}
\]
\[
H^{(y)}=\frac{1}{2}\sum_{i=1}^np_i^2+\sum_{i<j} \frac{(n_in_j+n'_in'_j+a_ib_j+b_ia_j)}{(h_i-h_j)(h_i^{-1}-h_j^{-1})}
\]
Here $n_i, n'_i$ can be expressed in terms of $N,N',H_i$ as in  (\ref{nN}), $p_1+\cdots +p_n=0$  and $h_1\cdots h_n=1$. Poisson brackets between 
$a_i,b_i,c_i$ are described above. Nonzero Poisson brackets between $p_i$ and $h_i$ are:
\[
\{p_i,h_i\}=\delta_{ij}h_i
\]

\subsubsection{The compact real form}
Consider now the real compact case $\g^*=su_n^*$. For the Poisson brackets between 
coordinate functions on $su_n^*$ we have:
\[
\{\mu_{kl},\mu_{k'l'}\}=i(\delta_{l,k'}\mu_{kl'}-\delta_{l'k}\mu_{k'l})
\]
where $\overline{\mu}_{ij}=\mu_{ji}$.

The rank one orbit has 
an open dense coordinate chart with 
\[
\mu_{ij}=\phi_i\overline{\phi_j}-\frac{\delta_{ij}}{n}(\phi,\overline{\phi})
\]
where $\psi_i=\overline{\phi_i}$ and $\psi_i'=\overline{\phi_i'}$. 
Poisson brackets between coordinates are

\[
\{\phi_k,\overline{\phi_j}\}=-i\delta_{kj}, \ \ \{\phi'_k,\overline{\phi'_j}\}=-i\delta_{kj}, \ \ \{\phi_k,\overline{\phi'_j}\}=\{\phi_k,\phi'_j\}=0
\]
As above, define
\[
N=\sum_i |\phi_i|^2, \ \ N'=\sum_i |\phi_i'|^2, \ \ n_i=|\phi_i|^2, \ \ n_i'=|\phi_i'|^2, \ \ H_j=n_j-n'_j
\]
and 
\[
a_i=\phi_i\overline{\phi_i'}, 
\]
Then, just as in a complex algebraic case we have 

\begin{equation}\label{fixedC}
|\phi_i|^2+|\phi_i'|^2=\frac{N+N'}{n}, \ \ |a_i|^2+\frac{H_i^2}{4}=\frac{(N+N')^2}{4n^2}, \ \
\sum_{j=1}^n H_j=N-N'
\end{equation}

Note that $b_i=\phi_i'\overline{\phi}_j=\overline{a}_i$ and we have the following Poisson brackets
\[
\{ a_k,\overline{a_j}\}=-i\delta_{kj} H_j, \ \  \{H_k, a_j\}=2i\delta_{kj} a_j
\]
Here we used $\{n_j,a_k\}=i\delta_{kj}a_k$. 

Therefore, in the real compact case we have an isomorphism
\[
(\cO\times \cO')\sslash H\simeq (\cO_0^{\times n})_{N-N'}
\]
where $\cO_0$ is a coadjoint orbit in $su_2^*$ with the Casimir element fixed being fixed as in (\ref{fixedC}). Topologically this orbit is a sphere. This can be seen easily in variables
\[
S^x_j=a_j+\overline{a}_j, \ \ S^y_j=i(a_j-\overline{a}_j, \ \ S^z_j=H_j
\]
For the Euclidean norm of $\bS_i=(S^x_i, S^y_i, S^z_i)$ we have
\[
(\bS_i,\bS_i)=4(|a_i|^2+\frac{H_i^2}{4})=(\frac{N+N'}{n})^2
\]
The quadratic Hamiltonians can be written as 
\begin{equation}\label{Hx}
H^{(x)}=\frac{1}{2}\sum_{i=1}^n(p_i+H_i+\frac{N-N'}{n})^2+\sum_{i<j} \frac{\varphi_h(\bS_i,\bS_j)+(\frac{N+N'}{n})^2}{(h_i-h_j)(h_i^{-1}-h_j^{-1})}
\end{equation}
\begin{equation}\label{Hy}
H^{(y)}=\frac{1}{2}\sum_{i=1}^n p_i^2+\sum_{i<j} \frac{(\bS_i,\bS_j)+(\frac{N+N'}{n})^2}{(h_i-h_j)(h_i^{-1}-h_j^{-1})}
\end{equation}
Here $\varphi_h$ is a rotation of vectors $\bS_i$ which acts on $a$-variables as
\[
\varphi_h(a_i)=a_ih_i
\]
and $h_j=\exp(iq_j)$ with $\{p_i,q_j\}=\delta_{ij}$.

The Hamiltonian (\ref{Hy}) describes $n$ identical particles with positions $q_i$ and momenta $p_i$ on a circle, each carrying classical spin variables described by vector $\bS_i$.

\section{Conclusion}

There are several interesting directions closely related to the subject and results of this paper.

One of these directions is the quantization of two--sided spin Calogero-Moser models. See \cite{ES} where trace functions were constructed as solutions to the dynamical KZ equation and the corresponding heat equation. These results 
give the quantization of the two--sided spin Calogero-Moser systems for the symmetric pair $G\subset G\times G$.
In the setting of split symmetric pairs the quantization can be regarded as the Harish-Chandra theory of $\tau$-spherical functions \cite{W}. In the quantum case, instead of coadjoint orbits $\cO_1$ and $\cO_2$,  one should
fix two finite dimensional $K$-modules $V_1$ and $V_2$. Quantum Hamiltonians
of such system are radial components of Casimir operators and their joint eigenfunctions
are $V_1\otimes V_2$-valued spherical functions. For details see \cite{SR}.

Another direction is the "nonlinear" deformation of models, where $T^*G$ is replaced by the 
Heisenberg double of the standard Poisson Lie group structure on $G$. For one--sided
model corresponding to the symmetric pair $G\subset G\times G$, such deformation is the relativistic 
spin Calogero-Model, see \cite{R2} and references therein. This model has a natural
interpretation in terms of moduli spaces of flat $G$-connections on a punctured torus.
It also has a natural generalization to moduli spaces of flat connections on surfaces \cite{AR}.
The relativistic version of the two--sided Calogero-Moser system corresponds 
to the moduli space of flat connections on a torus with two punctures. 
Quantization of these systems should give a version of Harish-Chandra theory for quantum groups.

Elliptic deformation of spin Calogero-Moser system was constructed
in \cite{KBBT} where integrability is established and solutions 
to equations of motion are constructed explicitly.  It is most likely that this elliptic deformation is only Liouville integrable but not superintegrable. 

Finally, we did not try here to describe symplectic leaves $\cS(\cO_1,\cO_2)$ as stratified symplectic leaves.
This is an interesting and important problem, which will be addressed in a separate publication.

\appendix

\section{Cotangent bundle \texorpdfstring{$T^*G$}{T* G} as a symplectic manifold}\label{cot}

First, let us fix some notations. The coadjoint action of the Lie algebra $\g$ on its dual vector space 
is defined as
\[
ad^*_x(l)(y)=-l(ad_x(y))
\]
where $l\in \g^*, x,y\in \g$ and $ad_x(y)=[x,y]$ is the adjoint action of the Lie algebra $\g$ on itself.

If $\{e_i\}$ is a basis in $\g$ and
\[
[e_i,e_j]=\sum_k C_{ij}^k e_k
\]
then
\[
ad_{e_i}^*(e^j)=-\sum_kC_{ik}^je^k
\]
where $\{e^i\}$ is the dual basis in $\g^*$,

The adjoint action of $\g$ on itself integrates to the adjoint action $Ad_g(x)$ of 
a corresponding Lie group $G$. The coadjoint action of the Lie algebra $\g$
on its dual vector space integrates to the coadjoint action $Ad^*_g(l)$ of the Lie group.

\subsection{$T^*G$ as a symplectic manifold}\label{tg}Here we assume that the cotangent bundle
is trivialized by right translations $T^*G\simeq \g^*\times G$, and we will describe explicitly 
Poisson brackets corresponding to the standard symplectic structure on this space.

It is easy to compute now the Poisson bracket between two smooth functions 
on $T^*G\simeq \g^*\times G$:
\begin{equation}\label{Pb}
\{f_1, f_2\}(x,g)=-(x,[\frac{\pa f_1}{\pa x},\frac{\pa f_2}{\pa x}])+(\pa^L_g(f_2), \frac{\pa f_1}{\pa x})-(\pa^L_g(f_1), \frac{\pa f_2}{\pa x})
\end{equation}
Here $(.,.)$ is the pairing between $\g^*$ and $\g$, $x\in \g^*$ and therefore $\frac{\pa f}{\pa x}\in \g$. The left gradient of a function on $G$ is
defined as $\pa^L_gf(g)\in \g^*$ with $(\pa^L_gf(g), X)=\frac{d}{dt}f(e^{tX}g)|_{t=0}$.

Sometimes it is convenient to have Poisson brackets written in terms of
coordinates and matrix element functions. Let $\pi$ be a representation of $G$ in a finite dimensional representation $V$. We will write
$\pi(g)$ for a linear map representing $g\in G$ and we will also think of $\pi(g)$
as a collection of matrix element functions $\pi(g)_{ab}$ in some basis $f_a\in V$. 

Poisson brackets between coordinate functions 
$l_i$ on $\g^*\subset T^*G\simeq \g^*\times G$ and matrix element functions $\pi(g)$ on $G\subset T^*G\simeq \g^*\times G$ are
\[
\{l_i,l_j\}=-\sum_kC^k_{ij} l_k, \ \ \{l_i, \pi(g)\}=\pi(e_i)\pi(g), \ \ \{\pi(g)_1, \pi(g)_2\}=0
\]
the last expression means that $\{\pi(g)_{ab}, \pi(g)_{cd}\}=0$ for all $a,b,c,d$.

\subsection{Left and right actions of $G$ on $T^*G$} The action of $G$ on itself
by left multiplication lifts to its action on $T^*G$ by left translations. 
After the trivialization of $T^*G$ by right translations the action of $G$ 
by left translations on $T^*G\simeq \g^*\times G$ is:
\begin{equation}\label{la}
h: (x, g) \mapsto (Ad_h^*(x), hg) .
\end{equation}

The vector field\footnote{Here we identify the vector field and the corresponding Lie derivative acting on functions} $(\delta^L_X f)(x, g) = \frac{d}{dt} f(Ad_{e^{tX}}^*(x), e^{tX}g)|_{t=0}$  on $T^*G$ generating
the left action of $X \in \g$ on $G$  is Hamiltonian:
\[
\delta^L_X f = \{H_X^L, f\}
\]
where $H_X^L = x(X)$, $X\in \g$, $x\in \g^*$. The corresponding moment map is
\[
\mu^L _G: T^*G \to \g^*, \qquad (x, g) \mapsto x.
\]

In terms of coordinates and matrix element functions $H^L_X=\sum_i X_il^i$
and 
\[
\delta^L_X l_i=-X^kl_jC^j_{ki}, \ \ \delta^L_X \pi(g)=\pi(X)\pi(g)
\]

\subsection{The right action of $G$ on $T^*G$.}
After the trivialization of $T^*G$ by right translations, the lift of the action of $G$ by right translations on $T^*G$ gives the following $G$-action on $\g^*\times  G$:
\begin{equation}\label{ra}
h: (x,g) \mapsto (x, gh\inv)
\end{equation}

The vector field $\delta^R_X f(x, g) = \frac{d}{dt} f(x, ge^{-tX})|_{t=0}$  on $T^* G$ generating the right action of $X \in \g$ on $G$  is Hamiltonian:
\[
\delta^R_X f = \{H_X^R, f\}
\]
where $H_X^R(x, g) = -(Ad^*_{g^{-1}}(x))(X)$.

The corresponding moment map is
\[
\mu^R_G : \g^* \times G \to \g^*, \qquad (x, g) \mapsto -Ad_{g\inv}^* x .
\]

In coordinates and matrix element functions $H^R_X=-\sum_i l_i (Ad_g(X))^i$ and
\[
\delta^R_X(l_i)=0, \ \ \delta^R_X(\pi(g))=-\pi(g)\pi(X)
\]

\subsection{The adjoint action of $G$ on  $T^*G$.}
After the trivialization of $T^*G$ by right translations, the adjoint action of
$G$ on itself lifts to the diagonal action of $G$ on $\g^*\times G$:
\[
h(x,g)=(Ad^*_h(x),hgh^{-1})
\]
Because the left and the right actions of $G$ on $T^*G$ are Hamiltonian,
the adjoint action is also Hamiltonian with 
\[
H_X(x,g)=H^L_X(x,g)+H^R_X(x,g)=x(X)-(Ad^*_{g^{-1}}(x))(X)
\]

The corresponding moment map is 
\[
\mu(x,g)=x-Ad^*_{g^{-1}}(x)
\]

\subsection{Quotient spaces}

It is clear that $T^*G/G\simeq G\backslash T^*G\simeq \g^*$ as a manifold 
and it is an easy exercise  to show that this is also an isomorphism of Poisson manifolds
where $\g^*$ is equipped with the Kirillov-Kostant Poisson structure (with minus sign).

Now let us describe the quotient space $T^*G/Ad_G$ for a simple Lie group $G$.

Fix Borel subalgebra $\mathfrak{b} \subset \g$ Denote by $\g_1^*=\oplus_{\alpha} \g_\alpha^*$ the subspace in $\g^*$ spanned by root subspaces.
We have $\g^*=\hh^*\oplus \g^*_1$ where $\hh^*$ is the dual space to the Cartan subalgebra $\hh\subset \mathfrak{b}\subset \g$.

The quotient space $(\g^*\times G_{reg}) / \Ad_G$ (in a sense of the GIT quotient) can be described as follows:

\begin{theorem}
The space $(\g^*\times G_{reg})/G$, as a Poisson variety, is  isomorphic to 
\[
(\g_1^*/ H \times T^* H_{reg}) / W
\]
where $\g_1^*/ H$ is the GIT quotient of $\g_1^*$ with respect to
the coadjoint action of the Cartan subgroup $H\subset G$.  The projection $\mu : \g^* \to \hh^*$ is the moment map 
for this action. The action of $H$ on $\g^*$ is Hamiltonian with respect to the Kirillov-Kostant 
Poisson structure. It defines the Poisson structure on $\g^*_1/H$. The factor $T^*H_{reg}$ has 
natural symplectic structure. 

\end{theorem}

\begin{proof}
Let $f_1$ and $f_2$ be functions on $T^* G$. Poisson brackets between 
these functions are given in (\ref{Pb}). Now assume that both these functions are 
$G$-invariant and $f_1$ depends only on $x$ and $f_2$ depends only on $g$.
Also, assume that $g$ is regular, $g=aha^{-1}$ where $h\in H_{reg}$, and denote
$\tilde{x}=a^{-1}xa$.
Then
\[
\{f_1,f_2\}(x,g)=(\pa^L_g(f_2), \frac{\pa f_1}{\pa x})(x,g)=(\pa_hf_2(h), \frac{\pa f_1}{\pa x}(\tilde{x}))
\]
In other words, this Poisson bracket is the standard Poisson bracket on the orbifold
$(\hh^*\times H_{reg})/W$.

From the same formula for Poisson brackets we can see that a zero weight polynomial 
in $\g_1^*$ Poisson commutes with functions on $\hh^*\times H_{reg}$. The ring of zero weight polynomials in $\g^*_1$ is the quotient subring of zero weight polynomials in $\g^*$, which completes the proof.

\end{proof}

\section{Poisson manifold $K\backslash T^* G/K$}\label{KG}

Let $K\subset G$ be a Lie subgroup. Because the actions of $G$ by left and right translations are Hamiltonian, such actions of its subgroup $K$ are Hamiltonian too
with the moment maps $\mu^{L,R}_K: T^*G\to \kf^*$:
\[
\mu^L_K(x,g)=\pi(x), \ \ \mu^R_K(x,g)=-Ad^*_{g^{-1}}(x)
\]
where $\pi: \g^*\to \kf^*$ is the projection map dual to the embedding map $\kf\to \g$. 

Spaces $T^*G/K$ and $K\backslash T^*G$ are Poisson manifolds. Their symplectic leaves
are given by the Hamiltonian reduction. 

\subsection{ The Poisson manifold $K\backslash T^* G/K$ }
Since actions by left and right translations are Hamiltonian, the double coset space 
$K\backslash T^* G/K$ is Poisson. Here we will describe an open dense subset 
of this coset when $G$ is simple and $K=G^\theta$ is the space of fixed points of the 
Chevalley automorphism.

We will say that $g\in G$ is regular if $g=k_1hk_2^{-1}$ and
$h\in H_{reg}$ and we will denote $G_{reg}$ as the subset of regular elements in. $G$. 

\begin{theorem} We have an isomorphism of Poisson varieties:
\[
K\backslash (\g^*\times G_{reg})/K\simeq (T^*H\times \kf^*\times \kf^*)/N_K(H_{reg})
\]
where the map is given by 
\[
K(x,g)K=K(\tilde{x}, h)K\mapsto ((\tilde{x}_0, h), \mu_K^L(x,g), \mu_K^R(x,g))
\]
The inverse map brings $(p,h), \mu, \mu')\in. T^*H\times \kf^*\times \kf^*$
to $(\tilde{x},h)\in \g^*, h)$ where $\tilde{x}_0=p$ and
\[
\tilde{x}_\alpha=\frac{\mu_\alpha'+h_\alpha\mu_\alpha}{h_\alpha-h_{-\alpha}}, \ \ \tilde{x}_{-\alpha}=\frac{\mu_\alpha'+h_{-\alpha}\mu_\alpha}{h_\alpha-h_{-\alpha}}
\]
\end{theorem}

\begin{proof}
Consider $G$ as a matrix group. Let $f_1,f_2$ be smooth functions on $T^*G\simeq \g^*\times G$. From (\ref{Pb}) we have the following expression for their Poisson brackets 
\begin{equation}\label{Pbh}
\{f_1,f_2\}(x,g)=tr(\frac{\pa f_1}{\pa x}[x,\frac{\pa f_2}{\pa x}])+tr(\frac{\pa f_1}{\pa x}g\frac{\pa f_2}{\pa g})-tr(\frac{\pa f_2}{\pa x}g\frac{\pa f_1}{\pa g})
\end{equation}
Assume that $f_1$ and $f_2$ are $K\times K$-invariant, $g\in G$ is regular and write $(x,g)=(k_1, k_2)(Ad_{k_1^{-1}}^*(x), h)$ where $g=k_1hk_2^{-1}$ and $h\in H_{reg}$,
$k_1,k_2\in K$ and $g=k_1hk_2^{-1}$. The pair is unique up to the action of $N_K(N_{reg})$. Denote $\widetilde{x}=Ad_{k_1^{-1}}^*(x)$. If $f_2$ is constant in contangent directions, i.e. if $f_2(x,g)=f_2(g)$, we have
\[
\{f_1,f_2\}(x,g)=tr(\frac{\pa f_1}{\pa x}(\widetilde{x}, h)h\frac{\pa f_2}{\pa h})
\]
Write $\widetilde{x}=\widetilde{x}_0+\sum_{\alpha\in \Delta} \widetilde{x}_\alpha e_\alpha$ where $\widetilde{x}_0\in \hh^*$ and $\widetilde{x}_\alpha$
are coordinates in the root basis $e_\alpha$. From (\ref{Pbh}) we have
\[
\{\alpha(\widetilde{x}_0), h_\lambda\}=\alpha(\lambda)h_\lambda, \ \  \{\widetilde{x}_\alpha, h_\lambda\}=0
\]
where $h_\lambda$ is the function on $H$ corresponding to weight $\lambda$.

Coordinate functions in $\widetilde{x}_\alpha$ can be expressed in terms of coordinate functions $\mu_\alpha, \mu_\alpha'$ from equations $\widetilde{x}=\mu$ and
$Ad^*_{h^{-1}}\widetilde{x}=\mu'$:
\[ \widetilde{x}_\alpha-\widetilde{x}_{-\alpha}=\mu_\alpha, \ \ -h_{-\alpha}\widetilde{x}_\alpha+h_\alpha \widetilde{x}_{-\alpha}=\mu'_\alpha
\]
Note that $\{\mu_\alpha, \mu'_\beta\}=0$ because the left and the right actions of $K$ on. $T^*G$ commute. Each set of variables $\mu_\alpha$ and $\mu'_\alpha$ satisfy the Kirillov-Kostant Poisson brackets by definition. For $\widetilde{x}_\alpha$ we have:
 \[
 x_\alpha = \frac{\mu_\alpha' + h_{\alpha} \mu_\alpha}{h_{\alpha} - h_{-\alpha}}, \ \ x_\alpha = \frac{\mu_\alpha' + h_{-\alpha} \mu_\alpha}{h_{\alpha} - h_{-\alpha}},
 \]
This proves the theorem.
\end{proof}

\section{Poisson manifold $G\backslash T^*(G\times G) /G$}\label{GG}

\subsection{The action of $diag(G)\subset G\times G$ on $T^*(G\times G)$ by left and right translations}

The action of $G \times G$ on $T^*(G\times G)\simeq \g^*\times \g^*\times G\times G$
by left translations
\[
(a,b): (x,y,g,h)\mapsto (Ad^*_a(x), Ad^*_b(y), ag,bh)
\]
is Hamiltonian with the moment map $\mu^L_{G \times G} (x, y, g, h) = (x, y)$. Therefore, the left action of the diagonal subgroup $G\subset G\times G$ is also Hamiltonian with  moment map 
\[
  \mu^L_G(x, y, g, h) = x + y.
\]

The action of $G \times G$ on $T^*(G\times G) \simeq \g^*\times \g^*\times G\times G$
by right translations $(g,h): (x,y,a,b)\mapsto (x, y, ag^{-1},bh^{-1})$ on is Hamiltonian with the moment map $\mu^R_{G\times G} : T^*(G\times G) \mapsto \g^* \times \g^*$:
\[
\mu^R_{G\times G} (x, y, g, h) = (-Ad^*_{g\inv} (x), -Ad_{h\inv}^* (y ))
\]
As a consequence the action of the diagonal subgroup $G\subset G\times G$ is also Hamiltonian with the moment map $\mu_G^R : T^*(G \times G) \to \g^*$:
\[
\mu_G^R(x, y, g, h) = -Ad^*_{g\inv} (x ) - Ad^*_{h\inv} (y).
\]

\subsection{Quotient spaces $G\backslash T^*(G\times G)$ and $T^*(G\times G)/G$}

Both quotient spaces $G\backslash T^*(G\times G)$ and $T^*(G\times G)/G$ are Poisson
manifolds. We have natural isomorphisms $T^*(G \times G) / G \xrightarrow{\sim} \g^* \times \g^* \times G$, 
\[
(x, y, g, h) \mapsto (x, y, gh^{-1})
\]
and $G\backslash T^*(G\times G)\simeq \g^* \times \g^* \times G$,
\[
(x, y, g, h) \mapsto (Ad^*_{g^{-1}}x, Ad^*_{h^{-1}}y, h^{-1}g)
\]

The action of $G$ by left translations on $T^*(G\times G)$ gives the action of $G$ on $\g^*\times \g^*\times G\simeq T^*(G\times G)/G$:
\[
h:(x,y,a)\mapsto (Ad^*_h(x), Ad_h^*(y), hah^{-1})
\] 
and the action of $G$ by right translations gives the following action of $G$ on $\g^*\times \g^*\times G\simeq G\backslash T^*(G\times G)$.
\[
h:(x,y,a)\mapsto (x,y, hah^{-1})
\]
In terms of the realization of cosets $T^*(G\times G)/G$ and $G\backslash T^*(G\times G)$ as the space $\g^*\times \g^*\times G$ the moment maps are:
\[
\mu^L(x,y,a)=x+y, \  \ \mu^R(x,y,a)=-x-Ad^*_a(y)
\]

Symplectic leaves of $G\backslash T^*(G\times G)$ and of $G\simeq T^*(G\times G)/G$
are given by the Hamiltonian reduction:
\[
\cS^L(\cO) = G\backslash (\mu_G^L)\inv(\cO)  = G\backslash \{(x, y, g, h) \mid x+y  \in \cO\} =G\backslash\{(x,y,a)| x+y\in \cO\}\subset G\backslash T^*(G\times G)
\]
\[
\cS^R(\cO) = G\backslash (\mu_G^L)\inv(\cO)  = G\backslash \{(x, y, g, h) \mid x+y  \in \cO\} =\{(x,y,a)| -x-Ad_a^*y\in \cO\}G\subset T^*(G\times G)/G
\]
where $\cO \subset \g^*$ is a coadjoint orbit, the $G$-actions and the moment maps are described above.

\subsection{The quotient space $G\backslash T^*(G\times G)/G$} 
Because actions of $G$ by left and right translations on $T^*(G\times G)$are Hamiltonian, the quotient space $G\backslash T^*(G\times G)/G$ is Poisson. 

Its symplectic leaves are given by the Hamiltonian reduction and also can be 
constructed as symplectic leaves of Poisson spaces $G\backslash \cS^R(\cO)$ or $\cS^L(\cO')/G$. 
\[
S(\cO,\cO')=\{(x,y,a|x+y\in \cO, \ \ -x-Ad_a^*(y)\in \cO'\}/G
\]
where the actions and moment maps are given above. Note that the automorphism $\theta$,
which permutes factors in $G\times G$ acts on $x,y,a$ as $x\mapsto y, y\mapsto x, a\mapsto a^{-1}$.

Now let us describe the Zariski open subset of regular elements $(\g^*\times \g^*\times G_{reg})/G\subset (\g^*\times \g^*\times G)/G\simeq G\backslash T^*(G\times G)/G$.

\begin{theorem}
We have the isomorphism of Poisson varieties
\[
(\g^*\times \g^*\times G_{reg})/G\simeq \bigl( (\g^* \times \g^*)/ / \Ad_H^* \times T^*H_{reg} \bigr) / W 
\]
where $(\g^*\times \g^*)/ /Ad^*_H$ is the Hamiltonian reduction with respect to 
the action of Cartan subgroup $H$\footnote{The Hamiltonian reduction is defined 
for Poisson manifolds as well. The result $\mu^{-1}(0)$ is another Poisson manifold.}. Geometrically, this is the GIT quotient. 
The isomorphism mapping is 
\[
G(x,y,a)\mapsto W(H(\mu^L_G(\widetilde{x},\widetilde{y}),\mu^R_G(\widetilde{x},\widetilde{y})), h)
\]
Here $a=bhb^{-1}$ for some $b\in G$ and $h\in H_{reg}$ abd $\widetilde{x}=b^{-1}xb$,
$\widetilde{y}=b^{-1}yb$. The inverse mapping brings $W(H(\mu,\mu'),h)$ to $G(\widetilde{x},\widetilde{y},h)$ 
\[
\widetilde{x_0}=\mu_0 + p, \ \ \widetilde{y}_0= - p, \qquad \widetilde{y}_\alpha = \frac{\mu_\alpha + \mu'_\alpha}{1-h_\alpha}, \qquad \widetilde{x}_\alpha = \frac{\mu_\alpha h_\alpha + \mu'_\alpha}{h_\alpha - 1}.
\]
Here we used  a root basis decomposition for $\widetilde{x},\widetilde{y},\mu,\mu'\in \g^*$, $\widetilde{x} = \widetilde{x}_0 + \sum_{\alpha \in \Delta} \widetilde{x}_\alpha e_\alpha$, $\widetilde{y} = \widetilde{y}_0 + \sum_{\alpha \in \Delta} \widetilde{y}_\alpha e_\alpha$, $\mu = \mu_0 + \sum_\alpha \mu_\alpha e_\alpha$, and $\mu' = \mu_0' + \sum_\alpha \mu_\alpha' e_\alpha$ with $x_0, y_0, \mu_0, \mu'_0 \in \hh^*$. We also assumed that $(\mu, \mu')$ are in the preimage for the moment map of the coadjoint diagonal $H$-action on $\g^*\times \g^*$, which results in $\mu_0=-\mu'_0$.
\end{theorem}

An outline of the proof. From the definition of the Poisson structure on $T^*(G\times G)$ we have the following formula for the Poisson brackets between coordinate and matrix element functions on $\g^*\times \g^*\times G\simeq T^*(G\times G)/G$.
\begin{align*}
\{f_1,f_2\}(x,y,a)=-(x,[\frac{\pa f_1}{\pa x},\frac{\pa f_2}{\pa x}])-(y,[\frac{\pa f_1}{\pa y},\frac{\pa f_2}{\pa y}]) \\+(\pa^L_a(f_2), \frac{\pa f_1}{\pa x})-(\pa^L_a(f_1), \frac{\pa f_2}{\pa x})-(\pa^R_a(f_2), \frac{\pa f_1}{\pa y})+(\pa^R_a(f_1), \frac{\pa f_2}{\pa y})
\end{align*}
Here $\pa^L_a$ and $\pa^R_a$ are left and right derivatives (as before)\footnote{This Poisson
structure can be written in terms of coordinates as follows. Let $x_i, y_i $ be coordinate functions on $\g^*$ corresponding to some basis $e_i$ in $\g$
and $\pi(a)$ be matrix element functions on $G$. We have:
\[
\{ x_i,x_j\}=-C_{ij}^kx_k, \ \ \{y_i,y_j\}=-C_{ij}^ky_k, \ \ \{x_i, \pi(a)\}=\pi(e_i)\pi(a), \ \ \{y_i, \pi(a)\}=-\pi(a)\pi(e_i)
\]}.

Assuming that $f_2$ does not depend on $x$ and $y$ we have:
\[
\{f_1,f_2\}(x,y,a)= (\pa^L_a(f_2), \frac{\pa f_1}{\pa x})-(\pa^R_a(f_2), \frac{\pa f_1}{\pa y})
\]
Now assume that $a=bhb^{-1}$ with $b\in G$ and $h\in H_{reg}$. Define $\widetilde{x}$ and $\widetilde{y}$ as above. Assuming that $G$ is a matrix group we can write
\begin{align*}
\{f_1,f_2\}(x,y,a)=tr(h \frac{\pa f_2}{\pa h}(h) \frac{\pa f_1}{\pa x}(\widetilde{x},\widetilde{y},h)) -tr( \frac{\pa f_2}{\pa h}(h)h \frac{\pa f_1}{\pa y}(\widetilde{x},\widetilde{y},h))
\end{align*}
From here it is clear that 
\[
\{\alpha(\widetilde{x}_0), h_\lambda\}=\alpha(\lambda)h_\lambda, \ \ \{\alpha(\widetilde{y}_0), h_\lambda\}=-\alpha(\lambda)h_\lambda
\]
Here $\alpha$ is a linear functional on $\hh^*$ and $h_\lambda$ is a function on $H$
corresponding to weight $\lambda\in \g^*$.

Thus, $\frac{1}{2}(x_0-y_0), h$ can be identified with a point on $T^*H_{reg}$ with its natural symplectic structure. Because the left and the right actions of $G$ on $T^*(G\times G)$ commute, Poisson brackets of functions of $\mu$ with functions of $\mu'$ vanish.
Poisson brackets of functions of $\mu$ with functions of $\mu$ are usual Kirillov-Kostant Poisson brackets, same for $\mu'$.

Conditions $x + y = \mu \in \cO$, $x + aya\inv \in \cO'$ imply:
\begin{alignat*}{2}
 \widetilde{x}_0 + \widetilde{y}_0 &= \mu_0, &\qquad \widetilde{x}_0 + \widetilde{y}_0 &= -\mu_0',\\
 \widetilde{x}_\alpha + \widetilde{y}_\alpha &= \mu_\alpha, &\qquad \widetilde{x}_\alpha + h_\alpha y_\alpha &= -\mu'_\alpha
\end{alignat*}
We can solve these equations for $\widetilde{x}_\alpha$ and $\widetilde{y}_\alpha$, assuming  $\mu_0 = - \mu_0'$, which results in formulae stated in the theorem. 

Symplectic manifolds $\cS_{reg}(\cO, \cO')$ are symplectic leaves of this Poisson manifold.

\section{Matrix element functions}\label{MEF}

\subsection{Poisson brackets for matrix element functions}
Fix a Killing form on a simple Lie group $\g$. This gives a linear isomorphism $\g^*\simeq \g$.
Let $\pi : G \to \End(V)$ be a finite dimensional representation. It also defines a finite dimensional representation of $\g$. Matrix elements $\pi_{ij}(g)$ in some basis $\{f_i\}$ are functions on $G$
and matrix elements $\pi_{ij}(x), x\in \g\simeq \g^*$ are functions on $\g^*$. 
Thus matrix elements $\pi_{ij}(x), \pi_{ij}(g)$ are functions on $T^*G\simeq \g^* \times G$. Let $\{e_a\}$ be a basis in $\g$ which is orthonormal with respect to the Killing form\footnote{The Killing form is given by the appropriately normalized trace on the adjoint representation $\langle x , y \rangle = \tr(ad_xad_y)$. Linear isomorphism $\g \simeq \g^*$ defined by  the Killing form identifies the adjoin representation and
the coadjoint representation. We assume that the basis is orthonormal, which means the structural coefficients $C_{abc}$ in $[e_a,e_b]=\sum_c C_{abc}e_c$ are totally skewsymmetric.} .
Denote 
\[
P_{12} = \sum_{a} \pi_1(e_a) \tensor \pi_2(e_a) \in \End(V_1) \tensor \End(V_2).
\]

The Poisson brackets between matrix elements functions with respect to the standard symplectic structure on $T^*G \simeq \g^* \times G$ (trivialized by right translations) can be written as:
\begin{alignat*}{2}
\{x_1, g_2\} &= P_{12} g_2, &\qquad \{x_1, x_2\} &= [P_{12}, x_2]\\
\{g_1, g_2\} &= 0.
\end{alignat*}
Here $x_1 = \pi_1(x) \tensor I$, $g_2 = I \tensor \pi_2(g)$ are functions on $T^* G$ with values in $\End(V_1 \tensor V_2)$.
In terms of coordinates $x_a$ on $\g^*$ corresponding to the basis $e_a$  in $\g$, these brackets are given in section \ref{tg}.

In terms of matrix element functions, left invariant vector field on $T^*G\simeq \g^*\times G$
can be written as 
\[
\delta^L_X \pi(x) = \pi([X, x]), \ \ \delta^L_X \pi(g) = \pi(X)\pi( g),
\]
Right invariant vector fields can be written as 
\[
\delta^R_X \pi(x) = 0, \ \ \delta^R_X \pi(g) = -\pi(g)\pi(X)
\] 
Both actions are Hamiltonian with Hamiltonians $H^L_X(x)=\tr(X x)$ and 
 $H_X^R(x, g) = -\tr(Xg\inv x g)$ respectively. Here the trace is taken over the adjoint representation.
 
Note that the matrix element functions $\widetilde x = g\inv x g$ on invariant with respect to the right $G$-action and 
\begin{align*}
\{\widetilde x_1, \widetilde x_2\} &= -[P_{12}, \widetilde x_2], \{\{\widetilde x_1, x_2\}=0\\
\{\widetilde x_1, g_2\} &= g_2 P_{12}.
\end{align*}
These identities are easy to verify:
\begin{align*}
\{\widetilde{x}_1,x_2\} & =\{g_1^{-1},x_2\}x_1g_1+g_1^{-1}\{x_1,x_2\}g_1+g_1^{-1}x_1\{g_1,x_2\} \\ &=-g_1^{-1}(-P_{12}g_1)g_1^{-1}x_1g_1-g_1^{-1}[P_{12},x_2]g_1+g_1^{-1}x_1(-P_{12}g_1)\\ & =g_1^{-1}P_{12}x_1g_1+g_1^{-1}x_1P_{12}g_1-g_1^{-1}P_{12}x_2g_1-g_1^{-1}x_1P_{12}g_1=0
\end{align*}
Here we used the identity $[P_{12}, x_1+x_2]=0$.

The Hamiltonian for the right $G$-action can be written as
$H_X^R(x, g) = -\tr(X \widetilde x)$.

\subsection{Poisson structure on  $T^*(G\times G)/G$ in terms of matrix element functions}

The coset space $T^*(G\times G)/G\simeq \g^*\times \g^*\times G$ is naturally a Poisson manifold. 
This Poisson structure can described in terms of matrix element functions $\pi(x), \pi(y), \pi(a)$.
Here $(x,y,g)\in \g\times \g\times G$ and we used the linear isomorphism $\g^*\simeq \g$ given by the 
Killing form. Nonzero Poisson brackets between these matrix elements functions are

\begin{align*}
	\{x_1, x_2\} &= [P_{12}, x_2] , &\qquad \{y_1, y_2\} &= [P_{12}, y_2]\\
	\{x_1, a_2\} &= P_{12}, a_2 , &\qquad \{y_1, a_2\} &= -a_2 P_{12}.
\end{align*}

Here are some other examples of Poisson brackets that are easy to compute using matrix element functions. 
Let $\widetilde y = aya\inv$, then we have:
\begin{align*}
\{\widetilde y_1, \widetilde y_2\} &= -[P_{12}, \widetilde y_2], &\qquad \{\widetilde y_1, y_2\} &= 0, &\qquad \{\widetilde y_1, \widetilde x_2\} &= -[P_{12}, \widetilde y_1]\\
\{\widetilde y_1, a_2\} &= -P_{12}, a_2, &\qquad \{x_1, \widetilde y_2\} &= [P_{12}, \widetilde y_2], &\qquad \{x_1 + \widetilde y_1, x_2 + \widetilde y_2\} &= [P_{12},x_2 + \widetilde y_2].
\end{align*}

Here is the proof of the last bracket:
\begin{align*}
 & \{x_1 + \widetilde y_1, x_2 + \widetilde y_2\} = [P_{12}, x_2] + [P_{12}, \widetilde y_2] - [P_{12}, \widetilde y_1] - [P_{12}, \widetilde y_2]\\
  &= [P_{12}, x_2] -  [P_{12}, \widetilde y_1] = [P_{12}, x_2] +  [\widetilde y_2, P_{12}] = [P_{12}, x_2 + \widetilde y_2]
\end{align*}
It also easy to see that  $\{x_1 + \widetilde y_1, a_2\} = 0$.


\begin{thebibliography}{99}

\bibitem{AR} S. Artamonov, N. Reshetikhin, {\em Superintegrable systems on moduli spaces of flat connections}, 
in progress.

\bibitem{Ca} F. Calogero, {\em Solution of the one-dimensional N-body problem with quadratic and/or inversely quadratic pair potentials}, J. Math. Phys. 12 (1971) 419-436.

\bibitem{CMbook} {\em Calogero--Moser--Sutherland Models}, Diejen, Jan F. van, Vinet, Luc (Eds.), CRM Series in Mathematical Physics, Springer, 2000.

\bibitem{ChF} O. Chalykh, M. Fairon, {\em On the Hamiltonian formulation of the trigonometric spin Ruijsenaars-Schneider system},  arXiv:1811.08727.

\bibitem{ES} Etingof P., Schiffmann O., {\em Twisted traces of quantum intertwiners and quan- tum dynamical R-matrices},  Communication in Mathematical Physics, 2001, Volume 218, n3, pp 633--663. math.QA/0003109.

\bibitem{FT} L. Feher, I. Tsutsui, {\it Regularization of Toda lattices by Hamiltonian reduction}, Journal of Geometry and Physics,  21(2):97-135 (1997), arXiv:hep-th/9511118.

\bibitem{Fe}  L. Feher, B.G. Pusztai, {\em Spin Calogero models associated with Riemannian symmetric
spaces of negative curvature} , Nucl. Phys. B 751 (2006) 436--458, math-ph/0604073.

\bibitem{FP1} L. Feher, B.G. Pusztai, {\em  A class of Calogero type reductions of free motion on a simple Lie group}, Letters in Mathematical Physics, 79(3), (2007), 263–277.

\bibitem{Fe1} L. Feher, {\em An application of the reduction method to Sutherland type many-body systems}, Geometric Methods in Physics. Birkhauser, Basel, 2013. 109–117.

\bibitem{GH} J. Gibbons, T. Hermsen, {\em A generalization of the Calogero-Moser system}, Physica 11D:337-348(1984).

\bibitem{CMspaces} P. Etingof and V. Ginzburg, {\em  Symplectic reflection algebras, Calogero-Moser space, and
deformed Harish-Chandra homomorphism}, Inventiones mathematicae
February 2002, Volume 147, Issue 2, pp 243--348, math.AG/0011114.

\bibitem{Ko} D. Kazhdan, B. Kostant and S. Sternberg, {\em Hamiltonian group actions and dynamical systems of Calogero type},  Comm. Pure Appl. Mathe 31:n4, 481-507(1978).

\bibitem{KLOZ1} S. Kharchev, A. Levin, M. Olshanetsky, A. Zotov, {\em Calogero-Sutherland systemwith two types interacting spins}, JETP 
Letters,  v. 106, (2017) No. 3 179–183; arXiv:1706.08793[math-ph].

\bibitem{KLOZ2} S. Kharchev, A. Levin, M. Olshanetsky, A. Zotov, {\em Quasi-compact Higgs 
bundles and Calogero–Sutherland systems with two types of spins}, J. 
Math. Phys., 59:10 (2018), 103509 , 36 pp., arXiv: 1712.08851.

\bibitem{Kod} L. Casian, Y. Kodama, {\em Toda lattice and toric varieties for real split semisimple Lie algebras}, In: {\em Integrable systems, Topology and Physics}, conference proceedings of a conference "Integrable systems in differential geometry",
Tokyo, 2000,  published in Contemporary Mathematics, by AMS. arXiv:math/9912021.

\bibitem{KBBT} I. Krichever, O. Babelon, E. Billey and M. Talon,{\em Spin Generalization of
Calogero-Moser system and the matrix KP equation},  hep-th/9411160.

\bibitem{LX} L.C. Li, P. Xu, {\em  Spin Calogero-Moser systems associated with simple Lie algebras}, C.R.Acad. Sci. Paris, Serie I, 331: n1, 55-61(2000); math.SG/0009180.

\bibitem{Mo} J. Moser, {\em  Three integrable Hamiltonian systems connected with isospectral deformations}, Adv. Math. 16 (1975) 197-220.

\bibitem{N}   Nekhoroshev, N.N., {\em  Action-angle variables and their generalizations}, Trans. Moscow Math. Soc. 26:180--197 (1972).

\bibitem{OP} M.A. Olshanetsky and A.M. Perelomov, {\em Classical integrable finite-dimensional systems related to Lie algebras}, Phys. Rept. 71 (1981) 313-400.

\bibitem{R} N.~Reshetikhin,
{\em Integrability of characteristic Hamiltonian systems on
simple Lie groups with standard Poisson Lie structure},  Comm. Math. Phys. {\bf 242} (2003), no. 1-2, 1--29, {\tt arXiv:math/0103147.}

\bibitem{R1} N.~Reshetikhin, {\em Degenerate integrability of the spin Calogero-Moser
systems and the duality with the spin Ruijsenaars systems},  Lett. Math.
Phys. 63 (2003), no. 1, 55--71.

\bibitem{R2} N.~Reshetikhin, {\em Degenerately Integrable Systems}, J Math Sci (2016)
213: 769.

\bibitem{STS} M. Semenov-Tian-Shansky :
\textit{Dressing transformations and Poisson group actions},
Publ.\ Res.\ Inst.\ Math.\ Sci.\ \textbf{21} (1985), 1237--1260.

\bibitem{SL} R. Sjamaar and E. Lerman, Stratified symplectic spaces and reduction, 
Ann. of Math. (2) 134 (1991), 375–422,

\bibitem{Suth}  B. Sutherland, {\em Exact results for a quantum many body problem in one dimension. II}, Phys.
Rev. A 5 (1972) 1372-1376.

\bibitem{SInt} {\em Superintegrability in Classical and Quantum Systems}, Edited by: P. Tempesta, P. Winternitz, J. Harnad, W. Miller, Jr., G. Pogosyan, M. Rodriguez, CRM Proceedings and Lecture Notes, Volume: 37, 2004.

\bibitem{SR} J. Stokman, N. Reshetikhin, {\it N-point spherical functions and asymptotic boundary KZB equations},
arXiv:2020.02251,

\bibitem{W} G. Warner, {\em Harmonic analysis on semi-simple Lie groups}, II, Springer Verlag, 1972. 

\end{thebibliography}
\end{document}